\documentclass[fleqn,usenatbib]{mnras}

\usepackage{newtxtext,newtxmath}

\usepackage[T1]{fontenc}
\usepackage{multirow}
\DeclareRobustCommand{\VAN}[3]{#2}
\let\VANthebibliography\thebibliography
\def\thebibliography{\DeclareRobustCommand{\VAN}[3]{##3}\VANthebibliography}

\usepackage{graphicx} 
\usepackage{wrapfig}
\usepackage{caption}
\usepackage{enumitem}
\usepackage{geometry}
\usepackage{blindtext}
\usepackage{multicol}
\usepackage{threeparttable}

\newcommand{\kms}{\,km\,s$^{-1}$\,}

\title[The coherent satellite velocity field around the interacting spiral galaxy pair NGC5713/19]{The coherent satellite velocity field around the interacting spiral galaxy pair NGC5713/19: signature of two galaxy groups merging}

\author[H. ~Jerjen et al.]{Helmut Jerjen$^{1}$\thanks{E-mail: helmut.jerjen@anu.edu.au}, Simon Deeley$^{2}$, Holger Baumgardt$^{2}$, Sarah M. Sweet$^{2,3}$
\\
$^{1}$Research School of Astronomy and Astrophysics, Australian National University, Canberra, ACT, 2611, Australia
\\
$^{2}$ School of Mathematics and Physics, University of Queensland, Brisbane, Qld 4072, Australia
\\
$^{3}$ ARC Centre of Excellence for All Sky Astrophysics in 3 Dimensions (ASTRO 3D)}

\date{Accepted 2025 July 24. Received 2025 July 24; in original form 2025 February 20}

\pubyear{2025}

\begin{document}
\label{firstpage}
\pagerange{\pageref{firstpage}--
\pageref{lastpage}}
\maketitle

\begin{abstract}
The luminous spirals  NGC5713 and NGC5719 form an interacting galaxy pair 94\,kpc apart and are connected by a  straight, elongated neutral hydrogen structure extending over 200\,kpc. Their 14 velocity-confirmed satellite galaxies and the two hosts separate into two distinct subgroups in their line-of-sight velocities and on-sky distribution revealing a prominent coherent kinematic structure with a velocity amplitude of $67\pm12$\kms. We test four scenarios to explain the observed velocity field: isotropic motions in a dark matter halo, a plane of satellites seen nearly face-on, and a kinematically mixed satellite system with a co-rotating edge-on plane and an isotropic component, and a merger of two small galaxy groups. Taking the geometry and dynamical state of the NGC5713/19 pair into account together with their positions and motions in the Bo{\"o}tes Strip the most consistent picture is the infall of two satellite systems that follow their host galaxies along a cosmic filament. We believe this is the first clear example of an equal-mass $L^*$ disk galaxy merger where a kinematically coherent satellite system is in the process of formation. These observations reinforce the importance of major mergers as a channel for producing co-rotating satellite systems.  
\end{abstract}

\section{Introduction}
\label{Introduction}
A main challenge for $\Lambda$CDM cosmology simulations on galaxy scales is posed by the plane-of-satellite phenomenon \citep{Bullock2017, DelPopolo2017, Pawlowski2018}. Observational evidence for dwarf galaxy satellite systems arranged in highly flattened planar structures, the so-called planes of satellites, is found for an increasingly large number of luminous ($L_K>10^{11}$\,L$_{K_\odot}$) host galaxies in the Local Volume ($D<10$\,Mpc), most notably the best studied cases of the Milky Way \citep{Lynden-Bell1976, Kroupa2005, Pawlowski2012, Pawlowski2015}, Andromeda \citep{Koch2006, Metz2007, Ibata2013} and Centaurus\,A \citep{Tully2015, Muller2016,Muller2018a, Muller2021}. Other Local Volume galaxy environments with flattened satellite planes include
M83 \citep{Muller2018, Muller2024},
NGC253 \citep{Martinez-Delgado2021},
M81 \citep{Chiboucas2013},
M101 \citep{Muller2017}, NGC2683 \citep{Crosby2023}, and NGC 4490/85 \cite{Karachentsev2024}. Possible exceptions are some of the 26 Milky Way-like hosts in the ELVES sample \citep{Carlsten2022} including the Sombrero galaxy M104 \citep{Crosby2023}. Outside the Local Volume, the first individual cases of flattened satellite structures around galaxies with stellar mass similar to the Milky Way ($L^*$ galaxies) have also emerged, such as NGC2750 at 40\,Mpc  \citep{Paudel2021}, while statistical studies confirm a high percentage of flattened satellite structures based on SDSS \citep[$>$50\%;][]{Ibata2014} and MATLAS \citep[$>$26\%;][]{Heesters2021,Heesters2024} Survey data. In the currently still few cases where velocities for a sufficiently large number of bright satellites around a host have been measured, i.e. Milky Way, Andromeda, Centaurus A, NGC4490/85, and NGC2750 these planar distributions were also found to form kinematically coherent structures supporting the notion that these satellite planes are co-rotating.

A key finding in the search for the origin of flattened, co-rotating satellite structures, and a matter of an on-going debate, is that they are rare in cosmological simulations, making them difficult to accommodate in the $\Lambda$CDM framework \citep[e.g.][]{Ibata2014, Bahl2014, Pawlowski2014, Pawlowski2020, Muller2021, Seo2024}. The processes put forward to explain such satellite configurations can be divided into three categories: preferential galaxy accretion along large-scale cosmic filaments \citep{Libeskind2010, Lovell2011, Shao2016, Mesa2018, Wang2020} such as the Local Sheet \citep{LocalSheet}, infall of a compact group of dwarf galaxies \citep{Metz2009, Julio2024}, and the formation of rotating tidal dwarf galaxy systems induced by host galaxy interaction \citep{Kroupa2005, Metz2007, Kroupa2012, Pawlowski2012, Hammer2013,Bilek2021}. For an in-depth review of the proposed solutions to the plane-of-satellite phenomenon, see \cite{Pawlowski2018}. 

One scenario that combines elements of all three processes is the close encounter and merger of two small galaxy groups that are moving along a cosmic filament, where each group comprises a single, luminous gas-rich $L^*$ galaxy with its own entourage of satellite galaxies. The most likely outcome would be a larger galaxy group with a more massive early-type host galaxy and a combined satellite system that also includes newly-formed tidal dwarfs. This is a probable future outcome for the Local Group and analogous to what we are currently witnessing in the NGC5128 group \citep{Israel1998}.
Hence, investigating close galaxy pairs and their environments in search for more clues as to how planes of satellites can form clearly is an important avenue to pursue. This has already been recognized and started. For example,
\cite{Libeskind2016} studied large numbers of galaxy pairs in SDSS and found that satellites preferentially occupy the space between the hosts with a 5-sigma significance. A subsequent theoretical study by \cite{Pawlowski2017} showed that pairs of host galaxies in cosmological simulations appear to be consistent with this particular empirical property of galaxy pairs. 

Intrigued by the recent findings of a phase-space-correlated population of satellite galaxies around the interacting Local Volume galaxy pair NGC 4490/85 \citep{Karachentsev2024,Pawlowski2024} and the interesting case of the NGC474/470 group that harbours a sub-population of six dwarf galaxies which seem to be anti-correlated in phase-space around the pair \citep{Muller2024} 
we investigated the galaxy pair NGC5713/19. This pair with its satellites turned out to be highly suited for the search for a co-rotating satellite plane for a number of reasons: (A) the pair is located outside the Local Volume at a distance of 27.2\,Mpc embedded in the Bo{\"o}tes Strip, a nearby cosmic structure, (B) the two spiral galaxies are in the process of merging with a projected separation between them of 94\,kpc, (C) deep 21\,cm mapping has revealed narrow bridges of neutral hydrogen gas connecting the galaxies and extending over 200\,kpc, (D) the linear HI distribution shows a well-defined line-of-sight velocity gradient that suggests the plane of interaction between the pair is seen nearly edge-on, (E) the host galaxies have comparable total masses, similar to MW and M31, resembling a future Local Group, (F) a sufficiently large number of 14 satellites have measured line-of-sight velocities to obtain a statistically robust result similar to the case of Centaurus A \citep{Muller2018}.

The paper is organized as follows. Section 2 discusses the cosmic environment and the unique properties of the NGC5713/19 pair. Section 3 introduces the satellite galaxies of the NGC5713/19 pair currently known within the group's virial radius, discusses their 2D distribution, and analyses the distinct velocity field. Section 4 presents various interpretations of the observed velocity field and in Section 5 we report the frequency and properties of NGC5713/19 analogues in the Illustris TNG100 simulations. Section 6 gives the summary and our conclusions.

\section{The galaxy pair NGC5713/19}
The galaxies NGC5713 and NGC5719 are located at a distance of $D=27.16$\,Mpc \citep[Cosmicflows-3;][]{Cosmicflows3}\footnote{https://vizier.cds.unistra.fr/viz-bin/VizieR-3?-source=J/AJ/152/50/table3} 
in the direction of (SGL, SGB)=(124.468\,deg, 25.324\,deg), which puts them 12.9\,Mpc above the Supergalactic Plane into the Bo{\"o}tes Strip \citep{Karachentsev2014}, a dispersed, large-scale cosmic filament of galaxies and galaxy groups behind the Local Void \citep{Kirshner1987} that connects to the Virgo Cluster. 
They make up the central galaxies of the GAMAJ1440-00 group \cite[GROUPID = 300150;][]{Robotham2011} with 15 Friend-of-friends group members and an estimated group radius of Rad100=314\,kpc\,h$^{-1}$. According to \cite{Karachentsev2014} and \cite{Kourkchi2017}  the NGC5713/19 pair is loosely associated with NGC5746 at a large projected distance of $\approx$1.2\,Mpc. 

\begin{figure*}
    \centering      
     \includegraphics[width = 17.5cm]{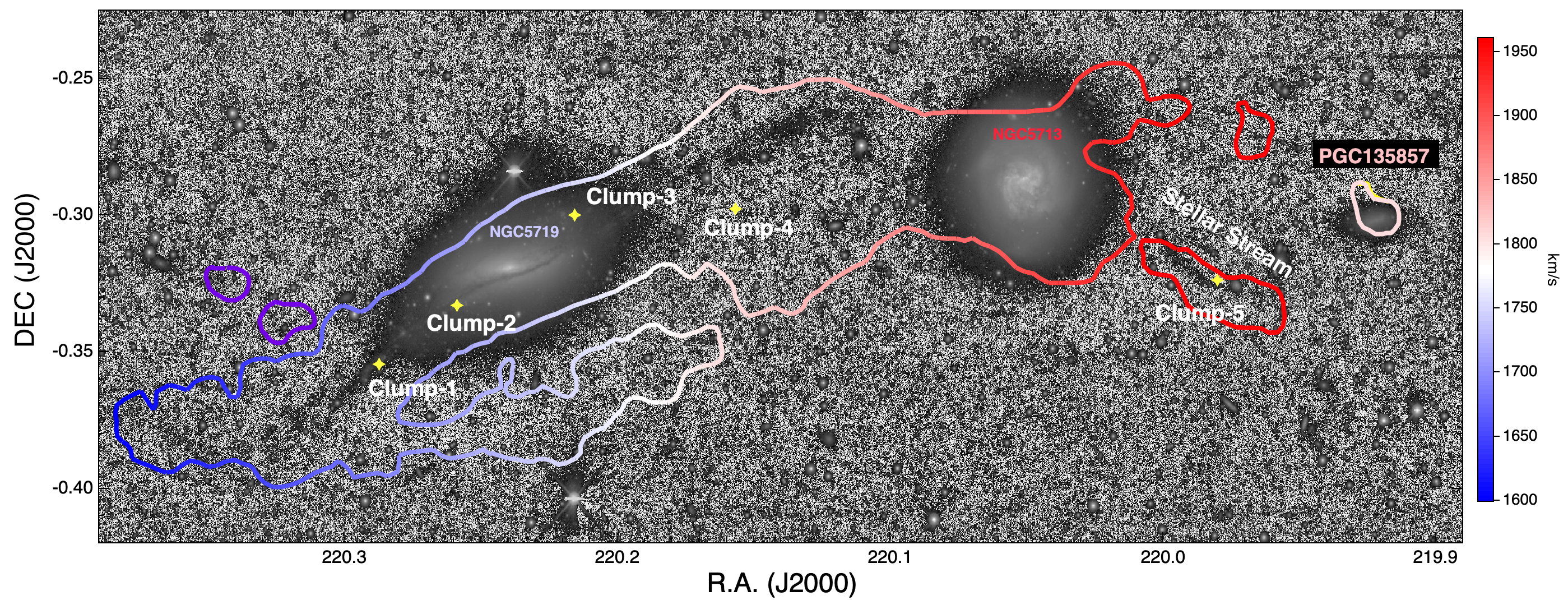}
    \caption{HSC-DR2 $g$-band image of the inner 200\,kpc around NGC5719 (left) and NGC5713 (right). Prominent stellar streams protrude on both sides of NGC5719, following the directions of the warped disk. Clump-1 and 2 are pockets of young stars forming in the stellar stream east of NGC5719 reported by \protect\cite{Neff2005}. The deep image also reveals a previously unknown stellar stream (R.A.$=219.98$, DEC$=-0.3228$) that coincides with the western HI tail that connects NGC5713 with the dwarf PGC135857, which most likely has emerged from the tidal debris of the interaction between the NGC 5713/19 pair. Three additional clumps (Clump-3 to 5) we could visually identify in the HSC-DR2 image.
    The contour corresponds to a HI column density of $7\times 10^{19}$ atoms \, cm$^{-2}$ reported by \protect\cite{Vergani2007}. The colours are an approximation of the velocity field of the HI filament stretching east-west and ranging from 1600 to 1960\kms (based on \protect\cite{Vergani2007}. The colour of the galaxy names reflect the measured LOS velocities.}
\label{galaxy_pair}
\end{figure*}

In Fig.\ref{galaxy_pair} we show the HSC-DR2 $g$-band image of the inner region of the GAMAJ1440-00 group. NGC5719 is the peculiar Sab galaxy to the east with a prominent warped disk seen almost edge-on. 
It is well known for its counter-rotating gas and stellar disks. The stars in the counter-rotating component have a significantly younger age of 0.7-2.0\,Gyr and lower metallicity \citep{Neff2005, Coccato2011} compared to the main stellar component (median age $\approx$4\,Gyr). These young stars are the result of retrograde accretion of external gas during the ongoing merger with NGC5713 \citep{Vergani2007, Coccato2011}. NGC5713 is the Sbc spiral located 94\,kpc to the west. It is oriented nearly face-on (inclination angle of the gas $i_{HI}=27\degr$ and stellar disk $i_{stars}\approx 10\degr$). This galaxy was a target in the $H_\alpha$ kinematics study by \cite{Urrejola-Mora2022} to 
measure its velocity field perpendicular to the disk. The authors concluded that the interactions of NGC5713 with both NGC5719 and the nearby dwarf PGC135857 are likely responsible for the lopsided morphology of NGC5713 and its unusual velocity field that is consistent with a vertically perturbed galactic disk.

\cite{Vergani2007} also fully mapped the neutral hydrogen gas around NGC5713/19. The VLA observations revealed a $\approx 200$\,kpc long, almost straight HI structure which commences at the south eastern side of NGC5719, connects the two  galaxies with two bridges and stretches all the way to the west of NGC5713. The intergalactic HI covers a velocity range from about 1610 to 1960\kms. In Fig.\ref{galaxy_pair} we overplot the HI contour 
that corresponds to a column density of $7\times 10^{19}$\,atoms\,cm$^{-2}$ taken from \cite{Vergani2007}. The colours along the contour represent the measured HI kinematics. A closer inspection of the deep HSC-DR2 $g$-band image and the comparison with the HI gas distribution shows that the formation of tidal dwarf galaxies is at work. Most prominently the dwarf galaxy PGC135857, which seems to have emerged from the tidal debris of the galaxy pair interaction \citep{Vergani2007}. On a more subtle level one notices stellar streams protruding on both sides of NGC5719's warped disk that are accompanied by diffuse UV emission and discrete bright blue pockets of young stars (Clump-1 and 2) reported by \cite{Neff2005} at the locations of the HI peaks. The locations of three additional clumps of young stars (Clumps 3 to 5) we visually detected are also labeled. The deep HSC-DR2 optical image further reveals a previously unknown stellar stream around Clump-5 that coincides with the western HI tail connecting NGC5713 with the dwarf PGC135857. 

We estimated the combined virial mass and radius for the NGC5713/19 pair as follows: the Galaxy Groups Within 3500\kms Catalogue \citep{Kourkchi2017}  provides apparent $K_s$-band magnitudes for both galaxies listed in Table\,\ref{pair}. These values are then converted to total $K_s$ luminosities adopting the distance of $D=27.16$\,Mpc ($m-M=32.17$) from \cite{Cosmicflows3} and using $M_{K_s, \odot}=3.27$ \citep{Willmer2018}. NGC5719 and NGC5713 have similar $K$-band luminosities, $7.2\times 10^{10} \,L_{\rm K_\odot}$ and $6.0\times 10^{10}\,L_{\rm K_\odot}$, respectively. The two galaxies also closely resemble Andromeda (SAb, $L_K=5.9\times 10^{10}\,L_{\rm K_\odot}$) and the Milky Way (SBbc, $L_K=4.7\times 10^{10}\,L_{\rm K_\odot}$) in terms of morphology \citep{RC3, Gerhard2002} and  $K$-band luminosity \citep{Kourkchi2017}.
Adopting a mass-to-light ratio of $\log(M/L_K)=-0.3$ for galaxies of that luminosity \citep{Bell2001} allows to calculate the individual stellar masses, which are converted to virial masses using the stellar mass - halo mass relation from \cite{Behroozi2013}. Values are listed in Table\,\ref{pair}. Total virial mass derivations yield similar estimates for NGC5713 (log(M$_{vir}/$M$_\odot)=12.22\pm0.24$) and NGC5719 (log(M$_{vir}/M_\odot)=12.15\pm0.20$). These masses are also comparable to the mass estimates for MW and M31 \citep{Klypin2002,Tully2015}. The combined total virial mass is thus log(M$_{vir}/M_\odot)=12.50\pm0.23$ which corresponds to a virial radius $R_{vir}=323^{+62}_{-52}$\,kpc and virial velocity $V_{vir}=205^{+40}_{-33}$\kms when adopting $H_0=67.5$\,km\,s$^{-1}$\,Mpc$^{-1}$ \citep{Planck}. Although these values for the group's virial radius and velocity are useful to have for reference they can only be rough estimates given the two host galaxies are in a pre-merging phase as we discuss in the following.\hfill

\begin{table} 
\begin{center}
\caption{Properties of the NGC\,5713/19 galaxy pair. Derived quantities are based on a distance of 27.16\,Mpc.}
\label{pair} 
\begin{tabular}{lll}
\hline
& NGC\,5719 & NGC\,5713 \\
& PGC52455  & PGC52412 \\
\hline
R.A. (J2000) & 14h40m56.4s &  14h40m11.4s \\
DEC (J2000)    & $-$00d19m06s &  $-$00d17m20s    \\
SGL & 124.467850$\degr$ &  124.374441$\degr$\\
SGB   &  25.323672$\degr$ & 25.154066$\degr$\\
Morph              & SAB(s)ab pec                & SAB(rs)bc pec               \\
$i_{HI}$     & 70$\degr$                            & 27$\degr$                           \\
P.A. (phot)      & 97\degr                             & 68$\degr$                         \\
$v_{\odot}$ & $1741\pm8$ \kms                     & $1899\pm8$ \kms                                    \\
K$_s$  & 8.29\,mag                       & 8.49\,mag                    \\
$M_{K_s}$  & $-23.88$\,mag                       & $-23.68$\,mag                    \\
$L_{\rm K}/ L_{\rm K_\odot}$         & $7.24\times 10^{10} $                      & $6.0\times 10^{10}  $                  \\
log(M$_*/M_\odot$)         & 10.56                       & 10.48                    \\
log(M$_{vir}/M_\odot$)         & $12.22\pm0.24$                       & $12.15\pm0.20$                    \\
V$_{vir}$        & $165\pm 34$ \kms                       & $157\pm 26$ \kms                   \\
\hline
\end{tabular}
\end{center}
\begin{tablenotes} 
\item NOTES: The centre positions are from NED, morphological type  are taken from the RC3 \citep{RC3}
The inclination,  position angle, and the heliocentric systemic velocities derived from HI data are from \cite{Vergani2007}. 
The total $K_s$ magnitudes are taken from \cite{Kourkchi2017}.
\end{tablenotes} 
\end{table}

\section{The Satellite System of NGC5713/19}
\begin{figure*}
    \centering    
     \includegraphics[width = 4cm]{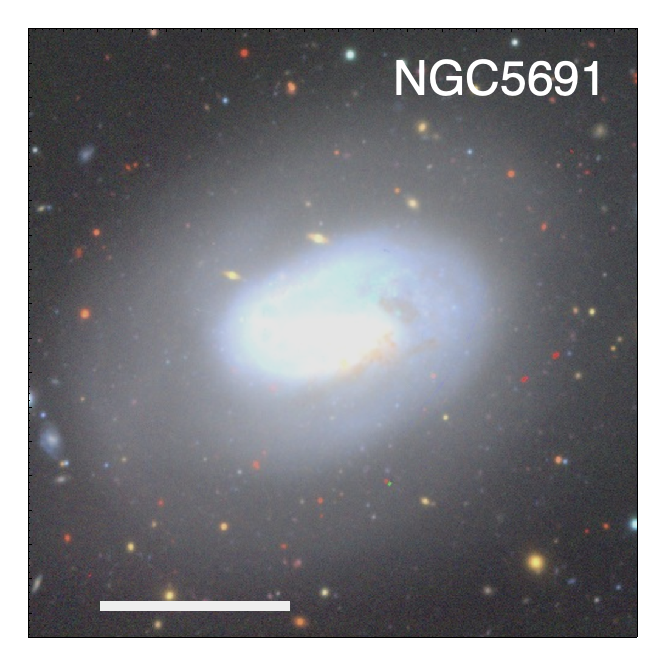}  
     \includegraphics[width = 4cm]{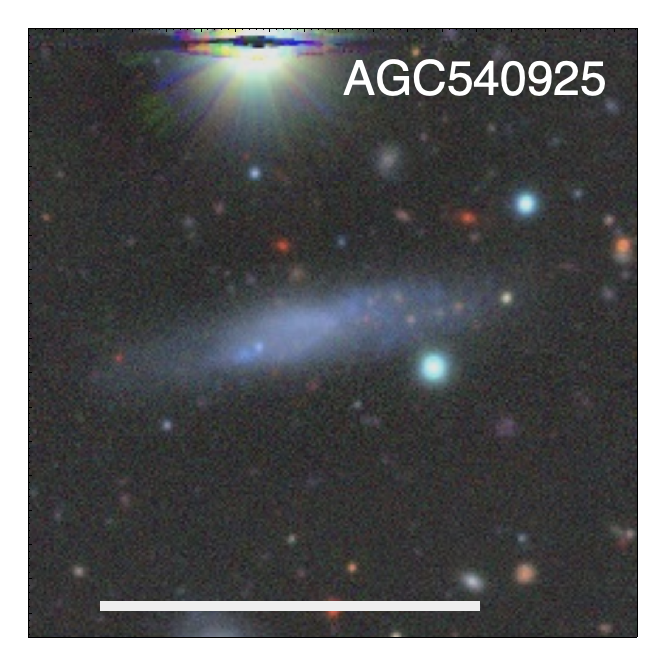}   
     \includegraphics[width = 4cm]{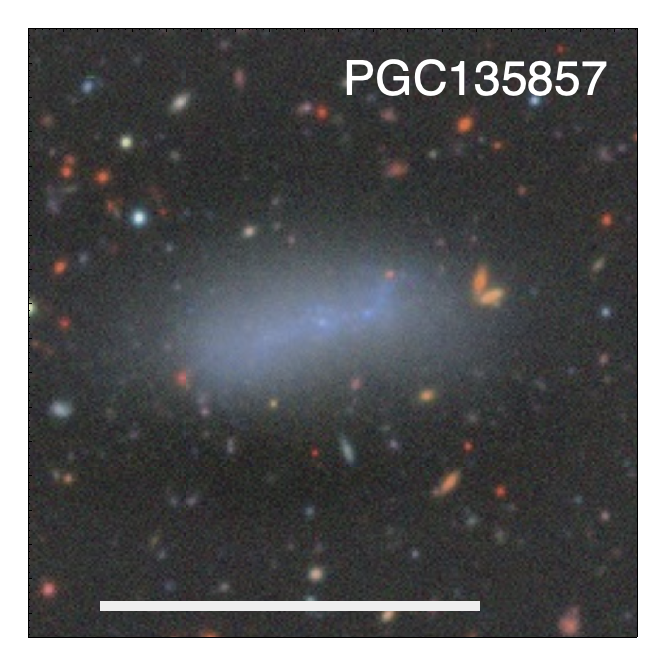}   
     \includegraphics[width = 4cm]{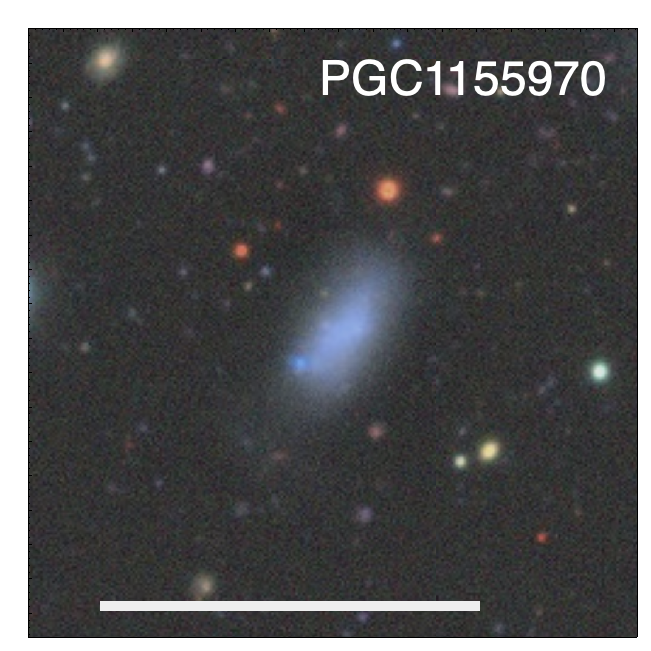}
     \includegraphics[width = 4cm]{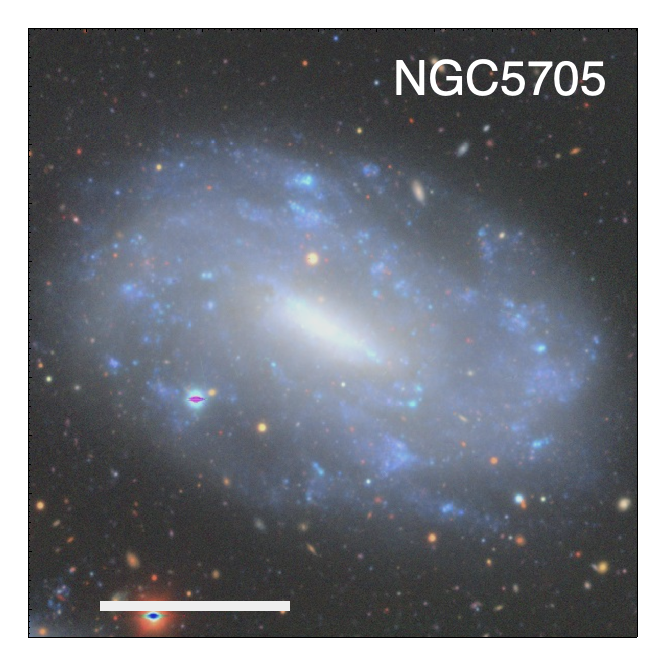}  
     \includegraphics[width = 4cm]{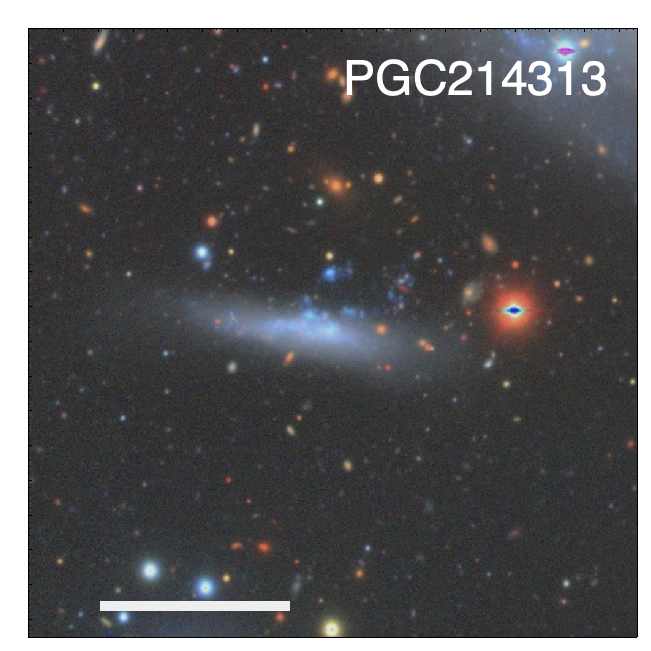}   
     \includegraphics[width = 4cm]{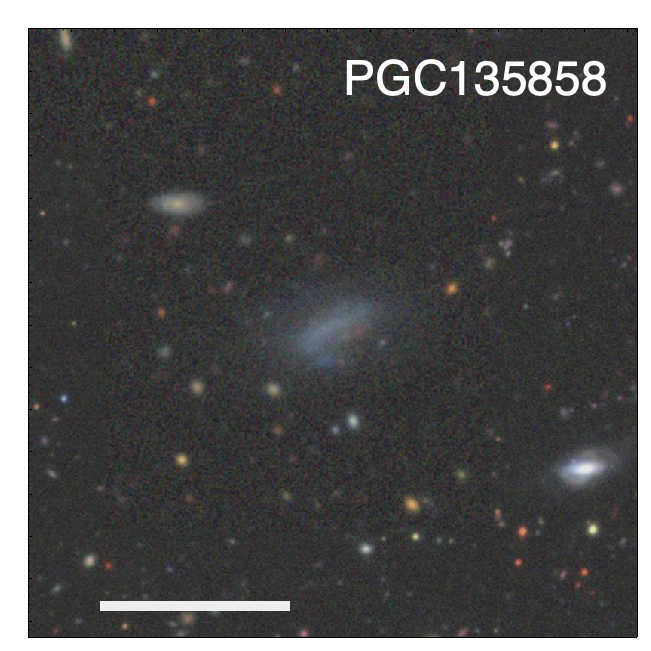}   
     \includegraphics[width = 4cm]{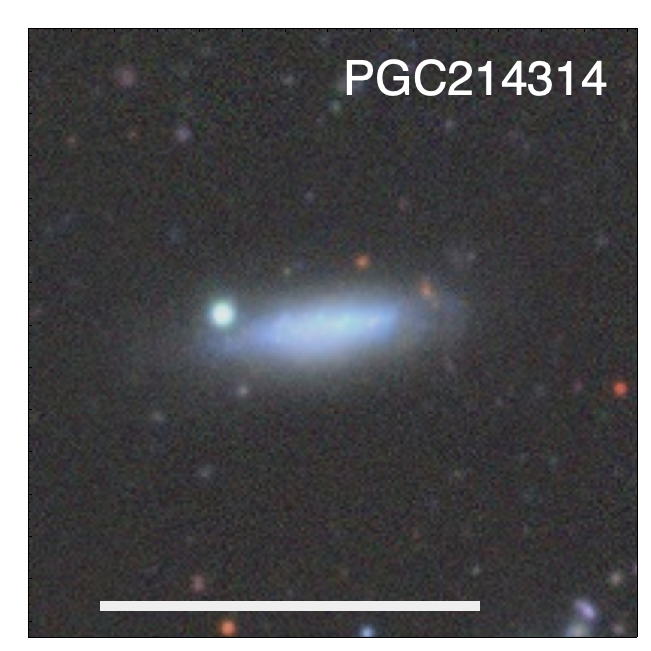}
    \includegraphics[width = 4cm]{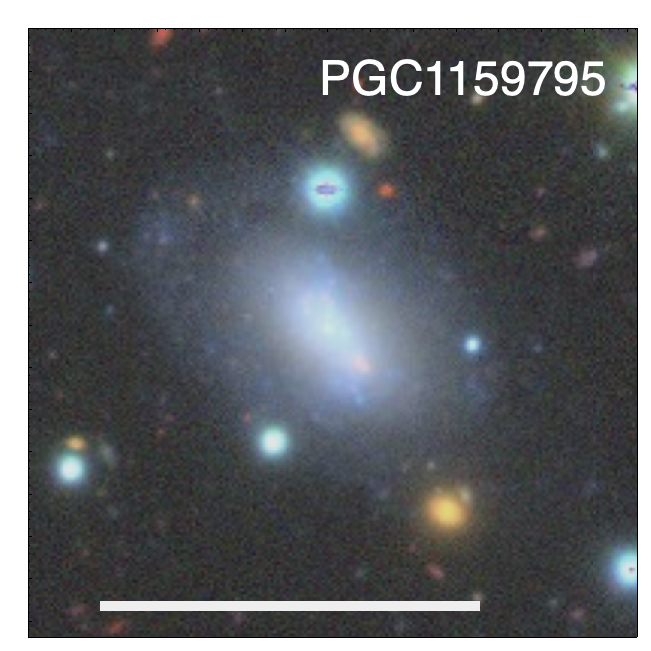}  
     \includegraphics[width = 4cm]{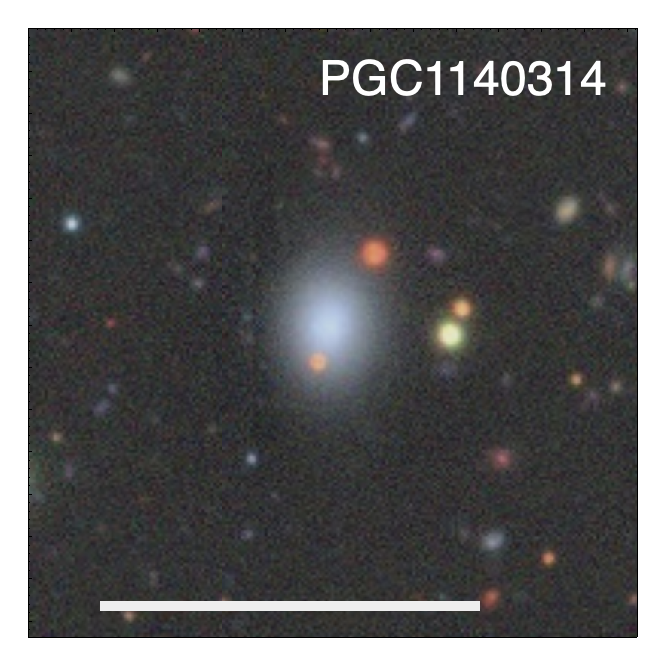}   
     \includegraphics[width = 4cm]{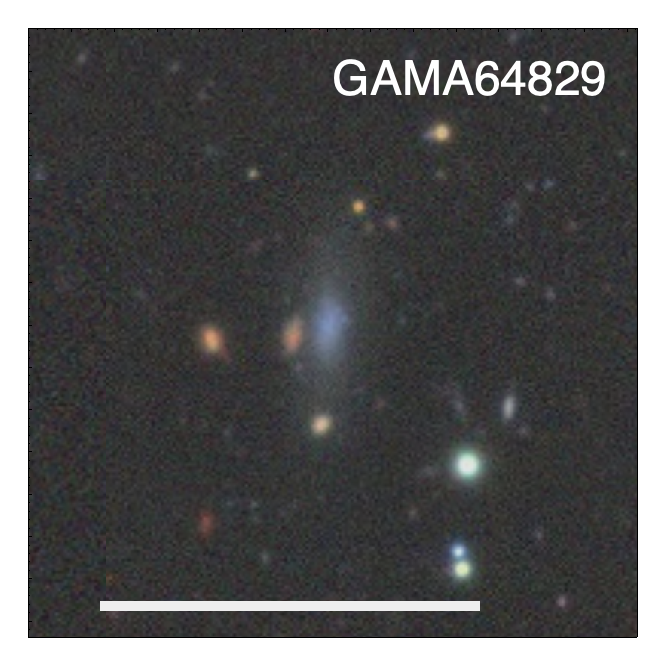}   
     \includegraphics[width = 4cm]{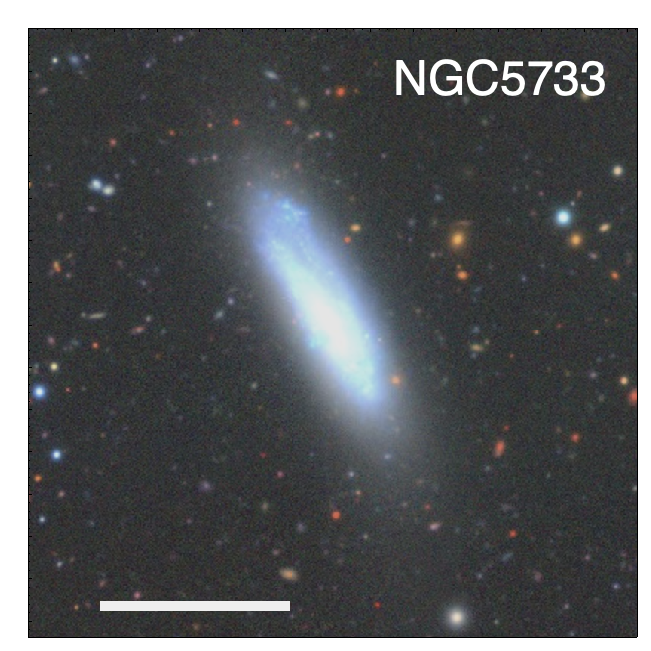}
    \includegraphics[width = 4cm]{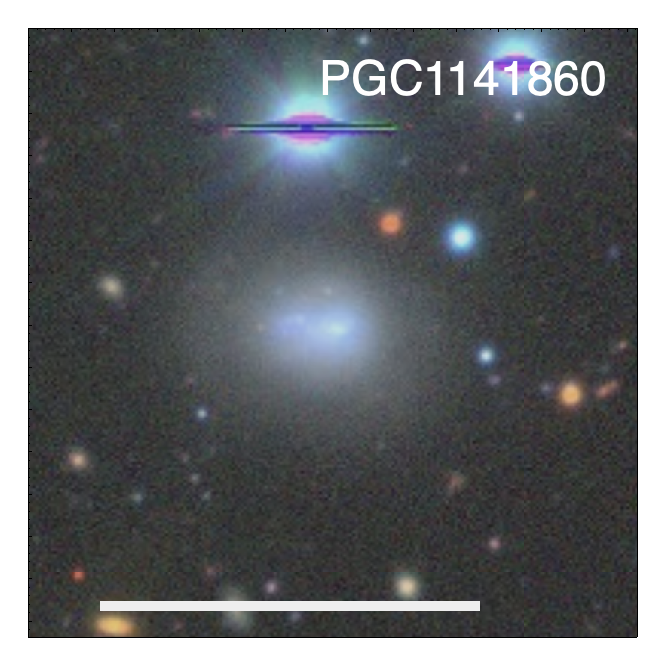}  
     \includegraphics[width = 4cm]{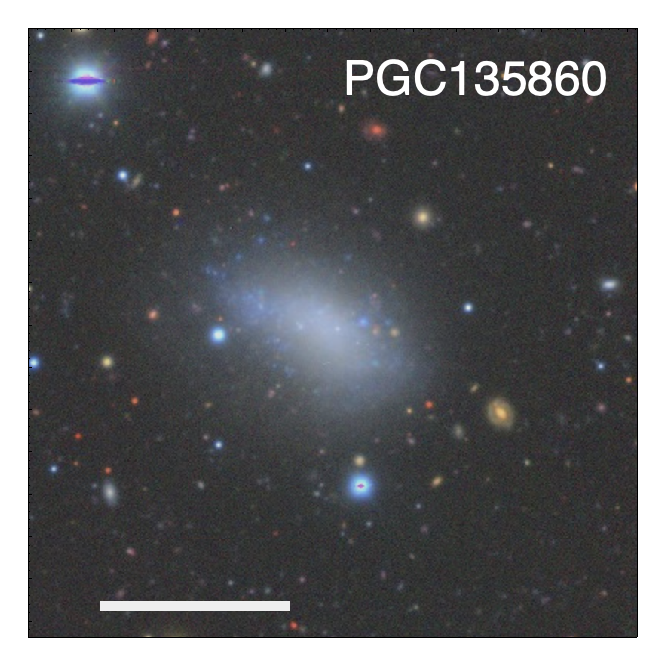}   
    \caption{HSC-SSP DR2 composite images of the 14 NGC5713/19 satellites that have measured velocities. From top left to bottom right they are sorted by R.A. to match the list in Table\,\ref{galaxyparameters}. All galaxies except PGC1140314 in third row have late-type morphology from Sc, Sd to irregular. Visually those galaxies appear to have at least a minimum level of ongoing star-formation and dust. The length of the horizontal bar is 30 arcsec, east to the left, north to the top.}
    \label{satellite_images}
\end{figure*}

\begin{figure}
    \centering    
          \includegraphics[width = 8.8cm]{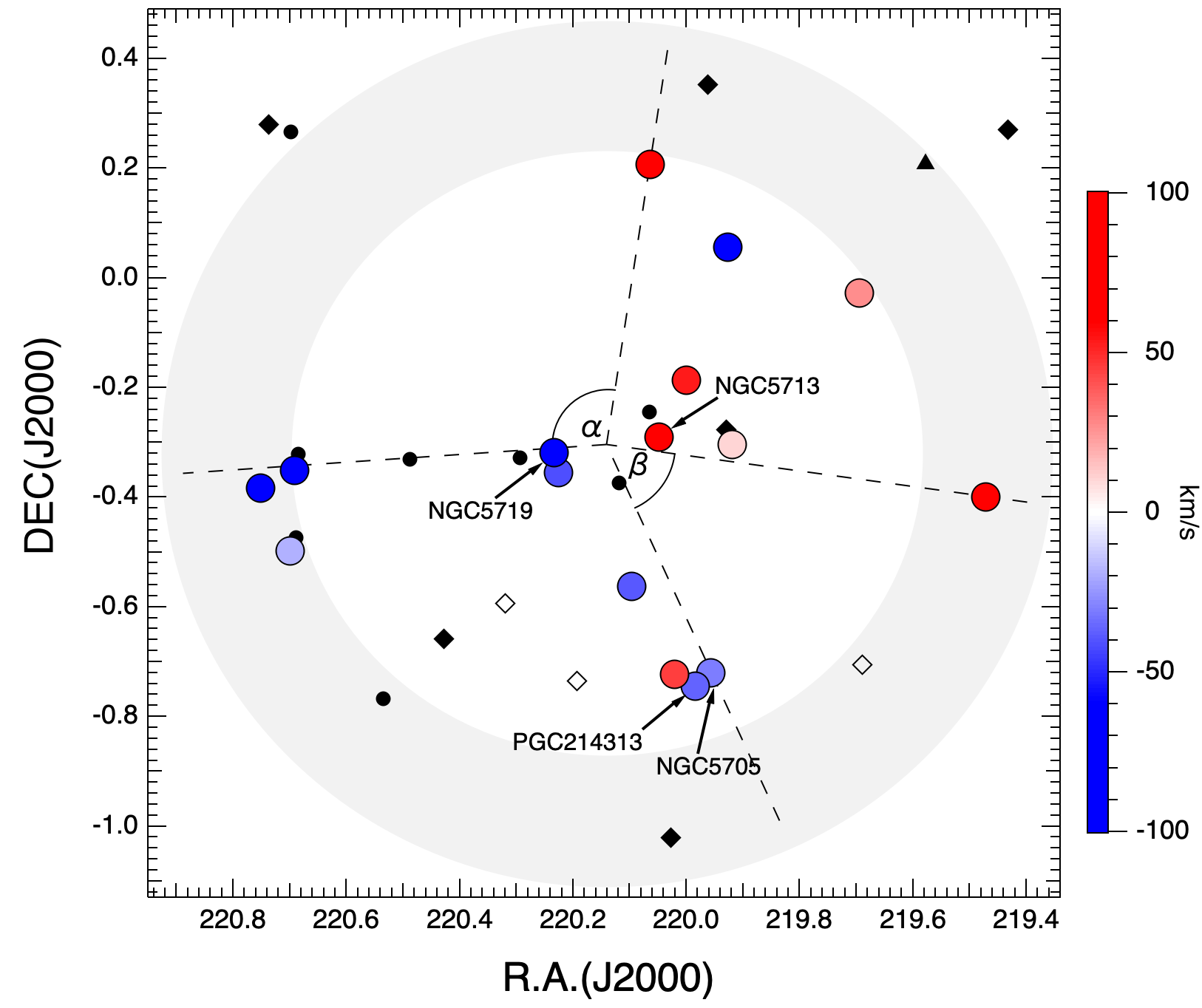}
    \caption{The on-sky distribution of the NGC5713/19 pair with its satellite galaxies. The satellites occupy two well-defined wedge-shaped regions within the maximum estimated virial radius of 385\,kpc (outer edge of grey annulus) centred at the midpoint between NGC5713 and NGC5719. Satellites with known velocities are color-coded by their velocity relative to the line-of-sight velocity 1805\kms (see Fig.\,\ref{vel_histogram}). All blueshifted galaxies except one are found in the wedge to the south east, including NGC5719 itself. To the north west are all but one redshifted satellites as well as NGC5713. This leaves two wide sectors with opening angles $\alpha=102.5\degr$ and $\beta=58\degr$, respectively, where no satellite with known velocity resides in. It is worth noticing that the bimodal distribution is also followed by GAMA79027 (triangle), four low-surface brightness satellite candidates (filled diamonds) reported by \protect\cite{Greco2018} and six newly detected dwarfs (filled circles) we report in this paper. Three more satellite candidates are just outside the maximum virial radius. The three satellite candidates from \protect\cite{Greco2018} shown as open diamonds have been reclassified as background galaxies by us based on the high-res HSC-DR3 images.}
    \label{onsky_distribution}
\end{figure}

A total of 14 satellite galaxies have measured velocities within the projected distance of 385\,kpc from the NGC5713/19 pair. Colour images from the Hyper Suprime-Cam Subaru Strategic Program Public Data Release 3 \citep{Aihara2022} are shown for all of them in Fig.\ref{satellite_images}. An additional 18 satellite candidates are known in the same area (listed in Table \ref{candidates}). They are dwarf galaxies with suspected group memberships based on  morphology and angular size but have no velocity or distance measurements to date: GAMA79027, nine low surface brightness (LSB) galaxies reported by \cite{Greco2018} and 
eight additional LSB galaxies. The latter sample we have visually identified through a meticulous inspection of the deep HSC-DR3 images following the approach outlined e.g. in \cite{Muller2015, Muller2017} covering the group area and using the images of the \cite{Greco2018} candidate as reference.
The on-sky distribution of all 32 galaxies is shown in Fig.\ref{onsky_distribution}. The satellites with velocities are highlighted in colour. Although the focus in this paper will be on the 14 satellites with velocities, we present the spatial information for the 18 satellite candidates in the appendix. A detailed analysis and discussion of the new candidates such as precise photometric properties will be provided in an upcoming paper.

From Fig.\,\ref{satellite_images} it is seen that all confirmed satellites of the NGC5713/19 group except PGC1140314 have late-type morphology from Sc, Sd to irregular. Visually all of them show evidence of ongoing star-formation.
Basic parameters for the satellites are listed in Table \ref{galaxyparameters}, including their GAMA IDs, total $K_S$-band magnitudes, and heliocentric velocities. References for the heliocentric velocities $v_\odot$ and uncertainties are given in column 10. For DESI velocities \citep{DESI2024} we adopted $\delta v_\odot=8$\kms for the uncertainty as per Fig.\,8 in \cite{Lan2023}. For the other galaxies we used the uncertainties quoted in the original sources. 

\begin{figure}
    \centering    
     \includegraphics[width = 8.7cm]{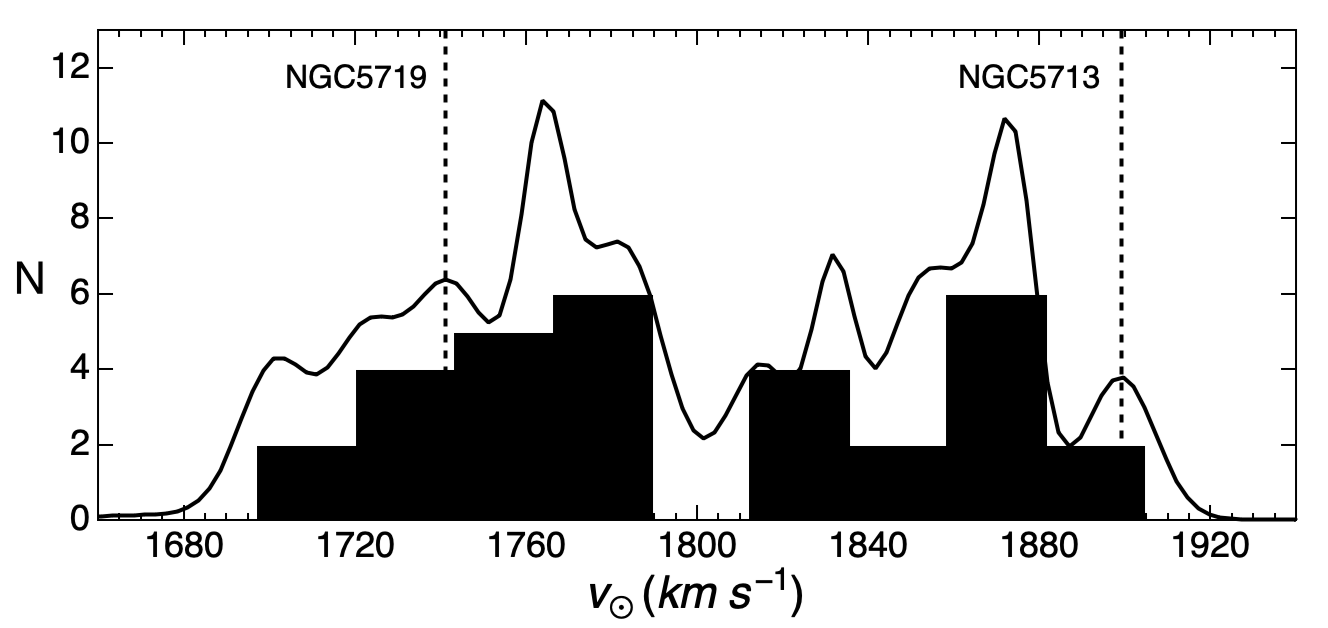}
    \caption{Histogram of the line-of-sight velocities for the 16 galaxies within the virial radius of NGC5713/19. The distribution shows a distinct bimodality with a gap between $1786-1815$\kms. The two vertical dashed lines indicate the velocities of the two host galaxies NGC5719 (1741\kms) and NGC5713 (1899\kms), respectively. For the solid line each galaxy is represented by a Gaussian profile centred at its measured velocity and the uncertainty used as the width. The line has a minimum in the gap of the histogram at 1800\kms.}
    \label{vel_histogram}
\end{figure}

In Fig.\,\ref{vel_histogram} we plot the raw histogram of line-of-sight velocities of the group members. Also shown is the corresponding kernel function where each galaxy velocity is replaced by a Gaussian profile with the velocity uncertainty taken as the standard deviation. Instead of a Normal line of sight distribution as expected for a system in dynamical equilibrium we find a distinct bimodal distribution with two groupings where each subgroup contains one of the two host galaxies, NGC5719 at 1741\kms and NGC5713 at 1899\kms, respectively. The gap without any galaxies is 29\kms wide ($1786< v_\odot <1815$\kms) and close to the mean systemic velocity of the host galaxy pair at 1820\kms. The kernel distribution has a significant dip in the same velocity range with a minimum at 1805\kms. This clear separation in line-of-sight velocity bears the question where are the satellite galaxies in the sky. This is shown in Fig.\,\ref{onsky_distribution} where galaxies are color-coded by their velocities relative to the line-of-sight velocity of $v_\odot =1805$\kms.

The two subgroups defined by their line-of-sight velocities also separate naturally in the sky (see Fig.\,\ref{onsky_distribution}). They occupy two well-defined sectors within the virial radius where the centre is at the midpoint between the NGC5713/19 pair. The satellite system shows an almost perfect correlation in phase space. All blueshifted satellites  except one (7 out of 8) are found in a wedge with opening angle 110\degr\,to the south east relative to the NGC5713/19 pair, including NGC5719 itself. In a sector to the north west we find all redshifted satellites except one (5 out of 6) as well as NGC5713. The latter sugbroup is confined to a wedge with opening angle of 89.5\degr. This leaves two sectors with opening angles $\alpha=102.5\degr$ and $\beta=58\degr$, respectively, where no galaxy with known velocity resides in. As the two sectors are almost opposite to each other
we propose a new metric, the zone-of-avoidance (ZOA) metric to test the significance of the distribution. This metric is  
 the maximum opening angle $\alpha_{ZOA}$ for a double wedge, two lines intersecting at a point, with opposite sides of the point considered as inside the double wedge, where no satellites are found. The concept is similar to the maximum opening angle metric used by \cite{Heesters2024} with the difference that the ZOA metric is testing for symmetry instead of lopsidedness. For the centre of the distribution we use the NGC5713/19 midpoint and we previously calculated $\alpha_{ZOA}=58$\,degree for the satellite distribution.
We determined the significance of the configuration by applying the ZOA metric to 100,000 Monte Carlo realisations, which were generated by assigning random position angles to the 14 satellites. In this way a $p$-value of 0.089 was found for the observed distribution, hence the level of separation and degree of symmetry are statistically significant at the $1.7\sigma$ level. If we include the two host galaxies the significance increases to $2.0\sigma$.

It is worth noticing that this bi-modal on-sky distribution prescribed by the confirmed group satellites is also followed by 11 additional
satellite galaxy candidates (see Fig.\,\ref{onsky_distribution}): 
GAMA79027 to the north west at the virial border, 
four low-surface brightness satellite candidates (filled diamonds) reported by \protect\cite{Greco2018} and six new dwarfs (filled circles) newly detected by us. Three more satellite candidates are just outside the virial radius. The three satellite candidates from \protect\cite{Greco2018} shown as open diamonds we consider to be background galaxies based on high-resolution HSC-DR3 images \citep{Aihara2022}. For further details about the satellite candidates see the Appendix.
We calculated the moment-based ellipticity parameters for the 14 velocity-confirmed satellites using their coordinates. The unweighted centroid of the distribution is at $({\rm RA}_{sat},{\rm DEC}_{sat})=(220.107\pm0.137\degr, -0.356\pm0.106\degr)$, the position angle of the minor axis $PA_{minor}=14.6\pm18.3$\degr (ccw, north), and the axis ratio $b/a=0.752\pm0.216$. Uncertainties are bootstrap estimates. The centroid offset is $d=222$\,arcsec or 30.1\,kpc away in projection from the nominal group centre, the midpoint of the NGC5713/19 pair at $({\rm RA}_{c},{\rm DEC}_{c})=(220.141 \degr, -0.305\degr)$. The value for the lopsidedness metric of normalised offset as defined by \cite{Heesters2024} was calculated, $d_{\rm norm}=0.139$ and calibrated. The obtained $p-$value 0.192 is small. Hence statistically there is no difference between the centroid of the satellite distribution and the position of the NGC5713/19 pair.

\subsection{Rotation Axis and Rotation Curve for the Satellites}
\begin{figure}
    \centering    
          \includegraphics[width = 8.8cm]{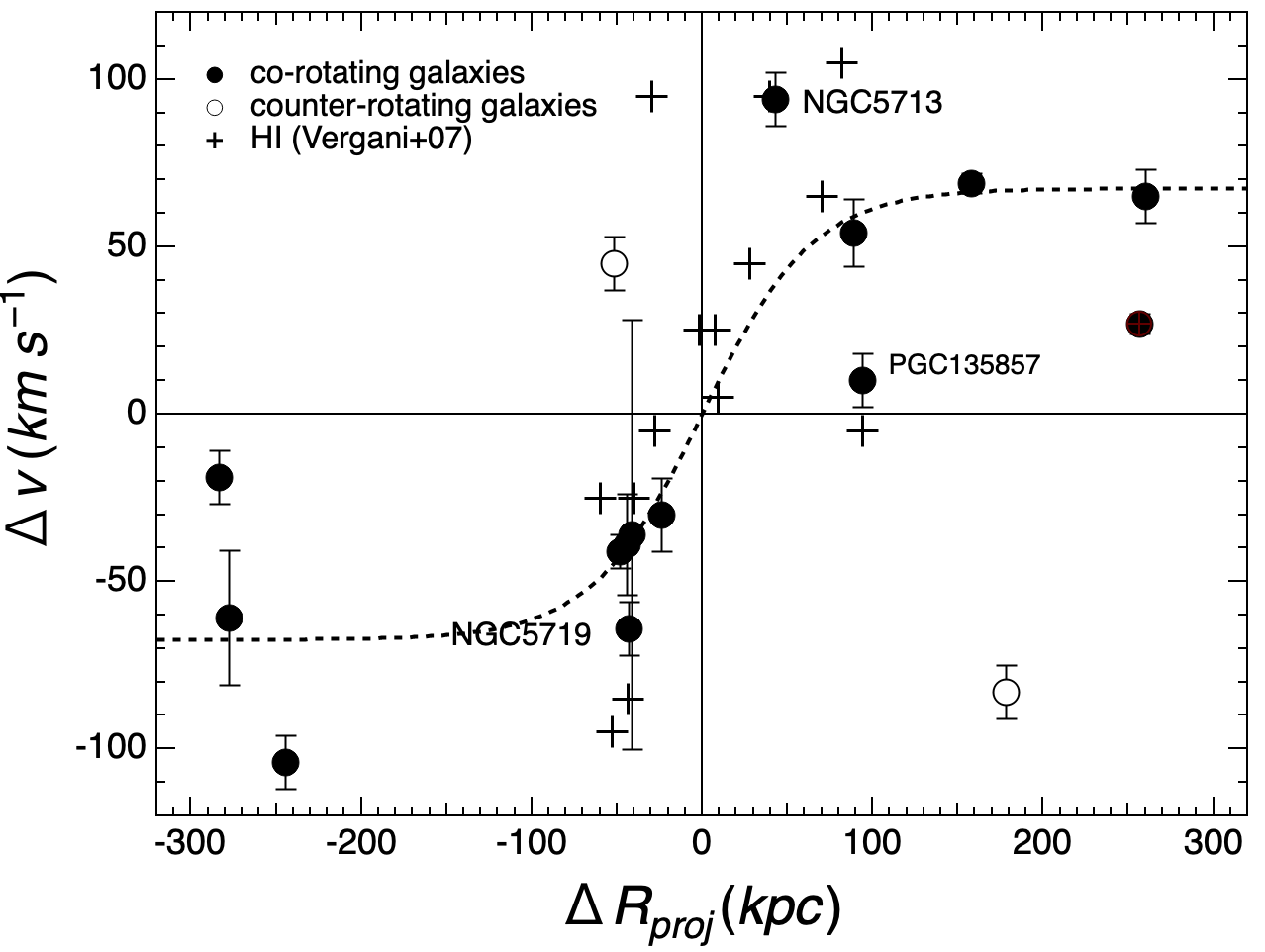}
    \caption{The position-velocity diagram of the NGC5713/19 group galaxies based on a projected rotation axis centred at the midpoint between NGC5713 and NGC5719 with a position angle  $PA=30\degr$ (ccw from north). Along the x-axis is the projected distance $\Delta R_{proj}$ to the rotation axis in kpc. The relative line-of-sight velocity is $\Delta v=v_\odot-1805$\kms. The galaxies show a phase-space correlation that resembles a flat rotation curve. We fitted a Sigmoid function to the data ignoring the host galaxies and the two counter-rotating satellites (open circles). The best-fitting function has an amplitude of $v_{max}=67 \pm 12$\kms. For comparison we also show data points (crosses) from the velocity field of the HI filament adopted from Fig.1 in \protect\cite{Vergani2007}. The tidal dwarf candidate PGC135857 is labeled.}
    \label{rotcurve}
\end{figure}

We found that the three observable phase-space parameters (RA,DEC,$v_\odot$) reveal a prominent, marginally flattened, co-rotating satellite galaxy structure, centred on the interacting NGC5713/19 pair and following the same sense of rotation. When determining the parameters for the most likely rotation axis and measuring the rotation curve for the satellite system we want to use the distinct velocity field in Fig.\,\ref{onsky_distribution} as a guidance. This requires the projected rotation axis to go somewhere between NGC5713 and 5719 with a position angle anywhere in the range $27\degr<PA<94\degr$ to keep it in the limits of the wedge with opening angle $\beta$. Furthermore, the systemic group velocity must be somewhere between $1786$\kms$<v_{sys}<1815$\kms.
Given the good agreement between satellite system centroid and midpoint between NGC5713 and NGC5719, we adopt the latter as crossing point for the rotation axis. For each combination of $PA$ and $v_{sys}$ values we calculated the relative velocities $\Delta v$ and the projected distances $\Delta R_{proj}$ from the rotation axis. The resulting rotation curve is shown in Fig.\ref{rotcurve}. Even without fine-tuning the parameters of the rotation axis, the axis offsets and relative velocities exhibit a S-shaped correlation 
that resembles a flat rotation curve. To maximise symmetry in the position-velocity diagram we fitted a Sigmoid function to the data ignoring the data from the two host galaxies and the two counter-rotating satellites (open circles). The best-fitting rotation curve is obtained for $PA=30\degr$ and $v_{sys}=1805$\kms, presented as a dashed line in Fig.\ref{rotcurve}. The maximum rotation velocity is $67\pm12$\kms, which is reached approximately at 100\,kpc radial distance. We note this value being significantly lower than the virial velocity of $V_{vir}=205^{+40}_{-33}$\kms derived in Section 2. This will be further investigated in Section\,4. The large velocity offset for PGC135857 can be explained by the possibility that this low mass dwarf located to the west of NGC5713 (see Fig.\ref{galaxy_pair}) is likely a tidally induced dwarf galaxy with a significant peculiar velocity relative to the systemic velocity. Both optical \citep{DESI2024} and 21cm \citep{Vergani2007} velocities for PGC135857 match and are  approximately $100$\kms lower than the nearby intergalactic gas ($v_{HI}\approx 1900$\kms).

An interesting comparison is provided by the known gas kinematics from the intergalactic HI that connects NGC5713/19 and stretches over 200\,kpc. For that purpose we overplot the position and velocity measurements of the data points in Fig.1 of \cite{Vergani2007} as crosses in Fig.\ref{rotcurve}. Although the HI was not used to establish the satellite system rotation curve, 11 out of 14 data points fall into the same quadrants as the galaxies (the other three are very close) and align well with the best-fitting rotation curve for the galaxies in the inner 50-100\,kpc. The linear increase in velocity as a function of distance has the signature of solid body rotation. This continuity of the HI in velocity can be used to estimate the age of the structure following \cite{Koribalski2004}. The velocity gradient along the main structure is (1960-1610)\kms/204\,kpc=1.715 \kms\,kpc$^{-1}$. The reciprocal value of the gradient gives a rough estimate for the time since the encounter of the two galaxies: $5.7\times 10^8$\,yrs, consistent with the age estimate for the young stars in the counter-rotating component of NGC5719 \citep{Neff2005, Coccato2011}.

The main difference between gas+host galaxies versus satellites is the significantly larger velocities reached by the hydrogen gas ($-195$ and $+155$\kms) and by the NGC5713/19 pair ($-64$ and $+94$\kms) approximately around the same radial distance of $\approx 50$\,kpc. Presumably angular momentum preservation has accelerated the galaxy pair with some of the surrounding HI gas to  approximately 2-3 times the velocities of the more distant satellite galaxies.

In Fig.\,\ref{plane_of_satellites} we show the best estimate for the projected rotation axis of the satellite system with $PA=30\degr$, centred on the NGC5713/19 midpoint. We also added an ellipse with an axis ratio $b/a=0.752$ representing the elliptical shape of the satellite distribution. 

\section{possible scenarios}
In order to test if the measured line-of-sight velocities of the satellite galaxies in the NGC5713/19 system are compatible with a spherical non-rotating distribution of galaxies, we set up an isotropic distribution of galaxies following an NFW density profile \citep{Navarro1996} and randomly draw repeatedly a sample of 14 galaxies from this system. We then determine the orientation on the sky that maximizes the measured rotation signal, i.e. the number of galaxies following a coherent rotation signal. We find that in only 11\,percent of the cases a rotation axis can be found so that at least 12 of the 14 galaxies follow a coherent rotation signal. If we include the two host galaxies and require that 14 out of 16 galaxies show rotation the probability drops to 4\, percent. Hence the coherent rotation signal that we find in the NGC5713/19 system is significant and unlikely to be due to a chance arrangement.

Then there are three alternative scenarios we can test the observed kinematically correlated satellite structure against to find the most likely origin of the observed velocity pattern: (i) a nearly face-on co-rotating plane of satellites, which would explain the observed low ellipticity of the satellite distribution and the low maximum line-of-sight rotation velocity, (ii) a nearly edge-on satellite system with a mixed (disk/isotropic) kinematics, which would explain the presence of some counter-rotating satellites, and (iii) an on-going merger between two host galaxies and their satellites. In the latter case the coherent velocity field  would be the superposition of group infall velocities and tidal motions in the newly forming galaxy group. We discuss each scenario in the following.

\subsection{Face-on co-rotating plane of satellites} 
\begin{figure}
    \centering    
          \includegraphics[width = 8.8cm]{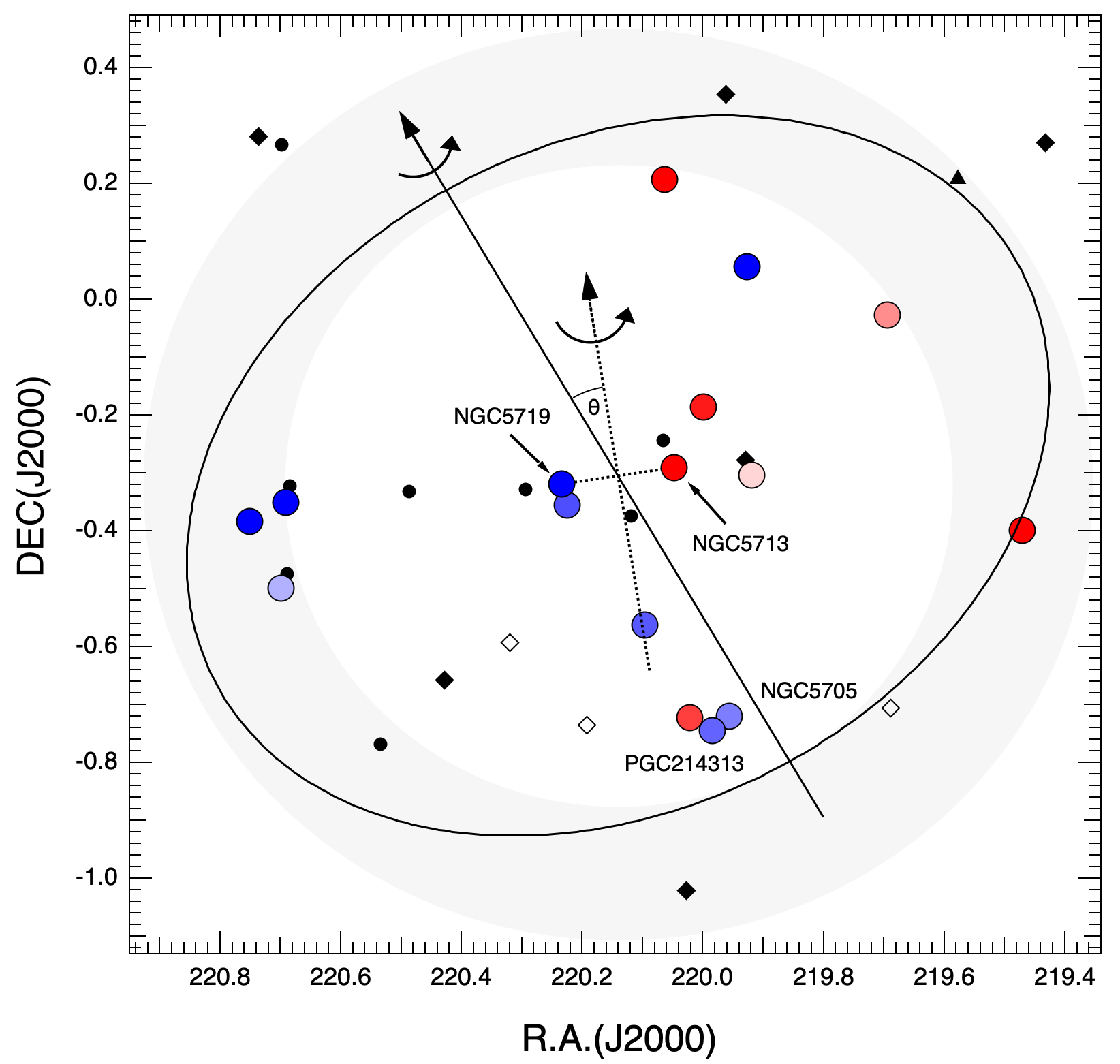}
    \caption{Modified version of Fig.\ref{onsky_distribution} to illustrate the best estimate for the projected rotation axis (solid arrow) 
of the satellite system around NGC5713/19.
    The rotation axis of the satellite system (solid arrow with direction of rotation) goes through the mid-point of the pair at a position angle of $PA=30^\circ$. That same line almost perfectly separates the satellites in terms of relative velocity to the systemic velocity of the group at 1805\kms: 7 out of 8 approaching on the NGC5719 side and 5 out of 6 receding on the NGC5713 side. We further note that the satellite rotation axis stands almost perpendicular to the NGC5713/19 connecting line (dotted line). If we assume the  plane of interaction between the two host galaxies is seen edge-on that means the rotation axis of NGC5713/19 (dotted arrow with direction of rotation) is in the plane of sky with a $PA=8.8$\degr. That gives $\theta=21.2$\degr for the angular separation between the two rotation axes. The ellipse with an axis ratio $b/a= 0.752$ is
representing the elliptical shape of the satellite distribution.}
    \label{plane_of_satellites}
\end{figure}

The 2D distribution of the satellites shows only moderate signs of flattening with an axis ratio $b/a=0.752\pm0.216$ (see Fig.\ref{plane_of_satellites}) while the observed maximum rotation velocity $v_{max}=67$\kms is significantly smaller than the virial velocity in the combined dark matter halo.

These two properties can be interpreted as an almost face-on flattened, co-rotating plane of satellites where the majority of galaxies on the eastern side move out of the plane-of-rotation towards us and most galaxies on the western side are moving away. The relative line-of-sight velocities would then be mainly the velocity component perpendicular to that plane. To de-project such a plane we assume an oblate shape with an intrinsic thickness of $q_0=0.2$, a value that lies between the derived values for the Milky Way ($q_0=c/a=0.209-0.301$) and M31 ($q_0=c/a=0.125$) planes of satellites \citep{Pawlowski2013}. From the observed axis ratio we can find the plane parameters using the standard equation:
\begin{eqnarray}
 \cos^2(i)= ((b/a)^2-q_0^2)/(1-q_0^2).
\end{eqnarray}
The satellite plane inclination is $i=42.3$\degr and the inclination-corrected rotation velocity $\bar{v}_{max}=99.6\pm 17.8$\kms. This corrected maximum velocity is still smaller than the virial velocity. The projected kinematic axis of the satellite plane would be almost parallel ($\theta=21.2$\degr) to the rotation axis of the NGC5713/19 interaction if assumed this to stand perpendicular to the connecting HI bridge as we will discuss in the next subsection.

In order to test if the line-of-sight velocity distribution of the dwarf galaxies is compatible with a rotating disc scenario, we set up a disc of satellite galaxies in an external NFW halo. We use the halo parameters for the NGC5713/19 group as we have determined them in Sec.~2 and assume a concentration parameter $c=8$ \citep{Ludlow2014} for the NFW halo. This gives us a halo scale radius of $R_S=40$\,kpc. For the disc of galaxies we assume an exponential disc with a scale radius of $R_d=30$ kpc and a thickness of $z_{d}=20$\,kpc independent of radius. We then set up 16 galaxies rotating with a circular velocity given by the NFW halo plus an additional random velocity component. We then view the disc under an inclination angle of $i=42$\degr  and calculate a Spearman rank order coefficient of the observed radial velocity of each disc galaxy against the projected distance of the galaxy from the center of the halo.

We find that the observed galaxies have a Spearman rank order coefficient of $r_s=0.46$, indicating a moderate degree of correlation between $\Delta R_{proj}$ and line-of-sight velocity relative to the mean velocity ($\Delta v_\odot$). For our modeled galaxies we find that for random velocities $\leq 0.30$ of the circular velocity of the NFW halo we usually obtain rank order coefficient of order $r_s=0.80$, significantly larger than observed. The rank order coefficients become comparable to the observed values only for random motions of order of the circular velocity. Hence the observed velocity distribution of galaxies would be compatible with a rotating disc scenario if the disc is kinematically hot.
 
\subsection{Satellite system with mixed kinematics} 
The scenario we discussed in the previous section 
took the phase-space properties of the satellites into account but ignored any additional information we have about the two central host galaxies that can be used to test another scenario: a satellite system with mixed kinematics, where the plane of satellites is seen almost edge-on.

\begin{figure}
    \centering    
          \includegraphics[width = 8.8cm]{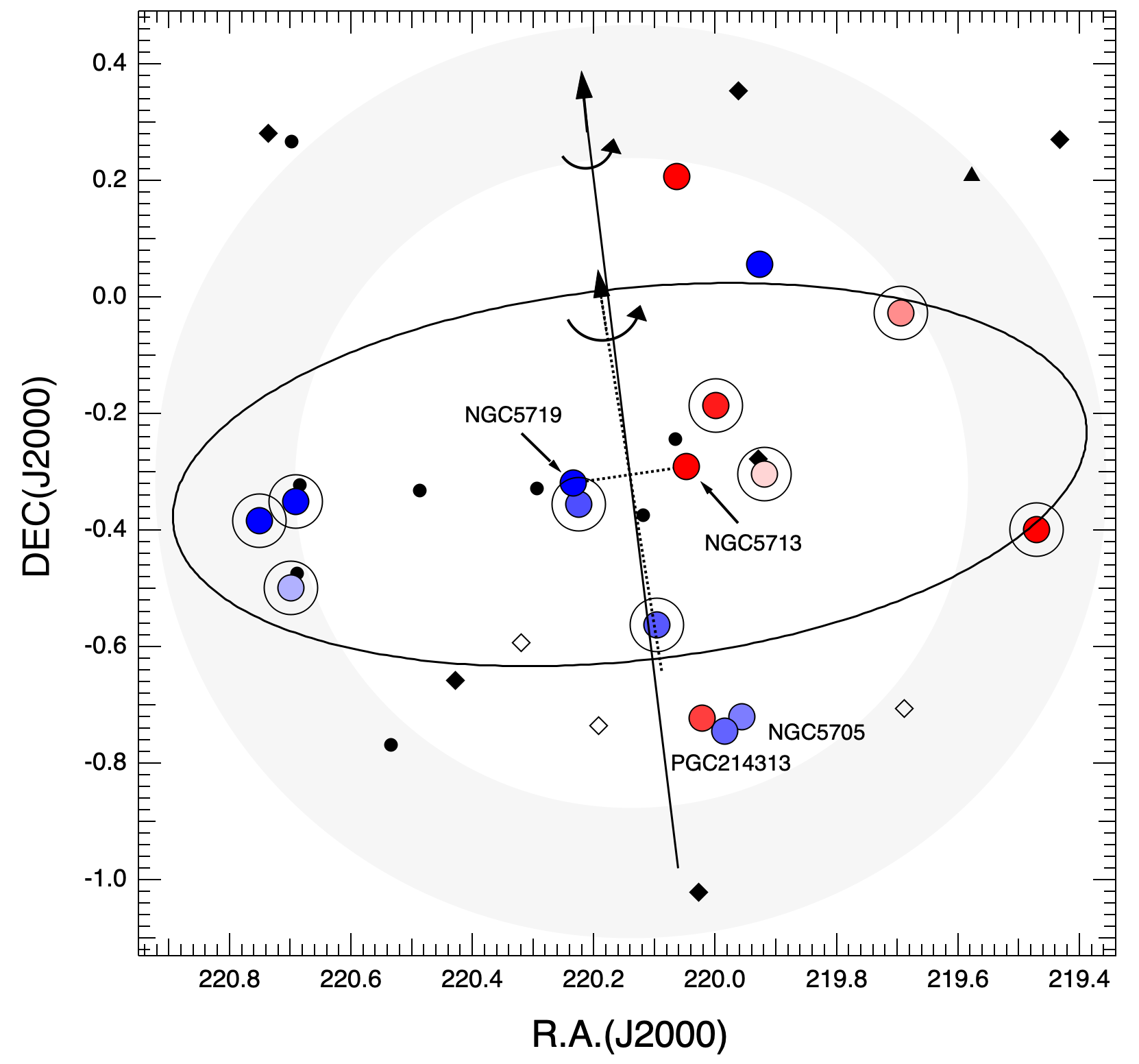}
    \caption{Modified version of Fig.\ref{onsky_distribution} to illustrate the mixed kinematics scenario. The nine coloured satellites highlighted with circles belong to the co-rotating plane (see also column 9 in Table \ref{galaxyparameters}) . Their rotation axis goes through the mid-point of the NGC5713/19 pair with a position angle of $6.7^\circ$. The ellipse with axis ratio $b/a=0.421$ represents the plane of rotation which is inclined by 68\degr relative to the line-of-sight. That means the satellite rotation axis is almost parallel to the rotation axis of NGC5713/19 (dashed line, PA=8.8\degr) and perpendicular to the HI bridge.  The narrow elongated east-west HI filament (Fig.\ref{galaxy_pair}) is represented by the dotted line connecting the central galaxy pair. The five coloured satellites without circles belong to the isotropic component. NGC5705 and PGC214313 are forming a gravitationally bound pair.}
    \label{edge_on_scenario}
\end{figure}

The central dotted line in Fig.\ref{edge_on_scenario} is not only the connection line between 
NGC5713 and NGC5719 but also 
a good approximation of the extended HI filament that connects the two host galaxies. It is a common feature of colliding equal-mass, gas-rich spiral galaxies in the early stages of merging that the HI distribution exhibits a complex morphology with prominent tidal arms when looking at the plane of interaction close to face-on \citep[e.g.][]{Holwerda2011, Sengupta2017}. In the case of NGC5713/19 the HI distribution has an elongated, straight viewing geometry with progressively increasing velocities from 1610-1960\kms (see Fig.\ref{galaxy_pair}) and no evidence of large tidal arms in the plane of sky. Furthermore, it has been known since  the early galaxy collision studies \citep{Toomre1972, Struck1999} that tidally induced stellar bridges tend to form and remain on two dimensional surfaces. These results support that the plane of interaction between the two host galaxies is seen almost edge-on ($i\approx90$\degr). In this case the rotation axis of the pair is in the plane of the sky and stands perpendicular to the connection line of NGC5713 and 5719, most likely going through the midpoint where the centre of mass is, given the two galaxies have similar total masses (see Table\,\ref{pair}). This rotation axis has a position angle of 8.8\degr and is shown as the dashed line in Figs.\,\ref{plane_of_satellites} and \ref{edge_on_scenario}. 

The majority of galaxies in the NGC5713/19 satellite system participate in co-rotation. However, the velocity field is not pure as we find one outlier on each side of the bimodal distribution. It is conceivable that the NGC5713/19 satellite system has a kinematically mixed satellite population like in the cases of M31 where 15 out of 27 satellites are in the Great Plane of Andromeda \citep{Ibata2013} and Centaurus A where 22 out of 27 satellites are in a plane \citep{Kanehisa2023a}. For an isotropic kinematic component we statistically expect an equal number of satellites that are blue and redshifted in each wedge. That means 4 out of the 14 satellites would belong to an isotropic component. So, we want to answer the question, "Which two co-rotating galaxies have to be allocated to the isotropic component to achieve the highest ellipticity in the remaining pure rotating component?".

Before doing this we want to establish whether or not the two most southern blueshifted galaxies NGC5705 and PGC214313 are a pair (see Fig.\,\ref{edge_on_scenario}). They are close in projection ($r_{proj}=18.1$\,kpc) and have a small line-of-sight velocity difference of $6\pm 65$\kms. We take the criterion from \cite{Geha2010}, where the lower limit on the required mass of a pair to be bound is $M_{pair}>r \Delta v^2/2 G$. As the spatial separation $r$ is unknown and we only have the line-of-sight velocity difference between the two we use $r=\sqrt{3/2} r_{proj}=22.2$\,kpc and $\Delta v=\sqrt 3 \times 71$\kms$=123$\kms. This gives $M_{pair}> 3.9\times 10^{10}$M$_{\odot}$. The combined total virial mass of the two galaxies can be independently calculated from their $K$-band luminosities (Table\,\ref{galaxyparameters}) by following the same procedure as used for NGC5713/19 before. We get $3.0\times 10^{11}$M$_{\odot}$, which is 7.7 times larger than the required minimum mass to be a gravitationally bound pair. So we treat the two galaxies NGC5705 and PGC214313 as one object in the following.

As there are six blueshifted independent satellites and five redshifted satellites to choose from on either side of the distribution there are 30 samples in total to work out ellipticity and position angle of the system. The highest ellipticity is achieved by allocating the NGC5705/PGC214313 pair and the most northern redshifted satellite PGC1159795 to the isotropic component.
The moment-based ellipticity parameters for the remaining clean, rotating satellite system of 9 galaxies are: $({\rm RA}_{sat},{\rm DEC}_{sat})=(220.151\degr, -0.378642\degr)$, the position angle of the minor axis $PA_{minor}=6.7$\degr (ccw, north), and the ellipticity is $1-b/a=0.579$. The centroid is $0.075$\degr (35\,kpc) away in projection from the nominal group centre.
 The lopsided metric of normalised offset is $d_{\rm norm}=0.168$, a value which indicates again no significant shift for the subsample of co-rotating satellites with respect to the hosts' centroid. The deprojection of the satellite plane as per equation (1) gives an inclination angle of $i_{plane}=68$\degr. 
 
 In summary we find that the rotation axis of the satellite plane  ($PA_{minor}=6.7$\degr) is almost parallel to the rotation axis of the NGC5713/19 interaction (dashed line, $PA=8.8$\degr) and perpendicular to the HI bridge. The other satellites without circles in Fig.\ref{edge_on_scenario} would belong to the isotropic component.  If this scenario is correct, the inclination-corrected maximum rotation velocity would be $72\pm13$\kms. 

\subsection{Ongoing merger of two galaxy groups} 
In Section\,2 we pointed out that the NGC5713/19 group is located in the Bo{\"o}tes Strip, a nearby cosmic filament behind the Local Void that connects to the Virgo cluster. At a closer look NGC5713/19 are part of an large-scale, elongated arrangement of Bo{\"o}tes Strip galaxies with similar radial velocities that stretches at almost constant declination ($DEC\approx0$\degr) from $R.A.=217.5$\degr to 221.5\degr. Moreover, the Bo{\"o}tes Strip has a systematic, shallow distance gradient between $212\degr<R.A.<228$\degr with a gradient approximately $-420$\,kpc/degree = $-860$\,kpc/Mpc at the location of NGC5713/19 as derived from Fig.2c in \cite{Karachentsev2014}, meaning that Bo{\"o}tes Strip galaxies west of NGC5713 are systematically closer to us than galaxies on the eastern side of NGC5719. This spatial distribution supports the picture where NGC5719 with its entourage of satellites moved in from the east and slightly behind the current location  while NGC5713 and its satellites approached from the west and front. The small projected distance between the galaxy pair (94\,kpc), the young kinematic age of the HI filament (0.6\,Gyr) and the most recent star formation episode observed in NGC5719 \citep{Coccato2011} are consistent with an early stage of merging. As a consequence of this scenario the satellites should still carry the signals from the infall velocity and direction while slowly getting unbound from their host halos and slung outwards to larger distances. The velocity of a satellite would be in first approximation the combination of the bulk flow motion of its group and its velocity induced by the tidal interaction as described in the disk-of-satellite formation scenario investigated by \cite{Smith2016}. In this context it is also important to point out the recent finding by \cite{Kanehisa2023b} that host galaxy mergers that efficiently transfer their angular momentum to satellite distributions are able to only marginally enhance their phase-space correlation, however cannot form highly flattened and orbitally coherent configurations as observed in the local Universe.

If the two host galaxies with their satellite systems are merging and the line-of-sight velocities of the NGC5719 satellites are  all blueshifted, then the line of sight component of the group infall velocity must be larger than the random velocity each satellite currently has on its orbit around the new centre of mass. The same train of thought applies for the NGC5713 group. In this scenario, the observed velocity map would not be the sign of a co-rotating plane of satellites in a dark matter halo, but the signature of two galaxy groups merging. 

We can get an idea how the galaxy distribution and motions along the line-of-sight  looks like by setting up a simple toy model. 
For that purpose we work in the phase-space ($\alpha$, $\delta$, $d$, $v_{\alpha}, v_{\delta},v_\odot$), assuming $\delta=0$, and $v_{\delta}=0$. Let the infall velocity vector for the NGC5719 group be $\vec{v}_{infall}^{NGC5719}=v_{infall}^{NGC5719}(\cos\theta, 0, \sin\theta)$, where $\theta$ is the infall direction ($\theta=0$\degr  being a transverse motion).
Furthermore, the velocity of a NGC5719 satellite around the new centre of mass shall be $\vec{v}_{tidal}^{NGC5719}=v_{tidal}^{NGC5719}(\cos\phi, 0, \sin\phi)$ where $\phi$ is the associated direction. If the line-of-sight normal vector is $ \vec{n}=(0,0,-1)$ then the heliocentric velocity can be written as:
\begin{eqnarray}
v_\odot=\langle (\vec{v}^{NGC5719}_{infall}+\vec{v}_{tidal}^{NGC5719}), \vec{n}\rangle\\
v_\odot=-
v_{infall}^{NGC5719} \sin\theta-v_{tidal}^{NGC5719}\sin\phi
\end{eqnarray}
where a negative sign for $v_\odot$ means blueshifted. We can now compare this equation with the observed rotation curve in Fig.\ref{rotcurve}. The first term on the RHS is the same for all NGC5719 satellites and thus can be approximated by the mean heliocentric velocity of the blueshifted part of the rotation curve: $\langle v_\odot\rangle =-67\pm12$\kms$=-v_{infall}^{NGC5719}\sin\theta$\kms. This can be solved for $v_{infall}^{NGC5719}$ by assuming the infall direction of the NGC5719 group is aligned with the distance gradient in the Bo{\"o}tes Strip, $\theta=-41\degr$. The  derived infall velocity then is $v_{infall}^{NGC5719}=-102\pm18$\kms. 
By the same train of thought we calculate $v_{infall}^{NGC5713}=102\pm18$\kms and $\theta=41\degr$ for the NGC5713 group. The second term on the RHS, $v_{tidal}^{NGC5719}\sin\phi$ is the offset from the mean velocity.

\begin{figure}
    \centering    
          \includegraphics[width = 8.8cm]{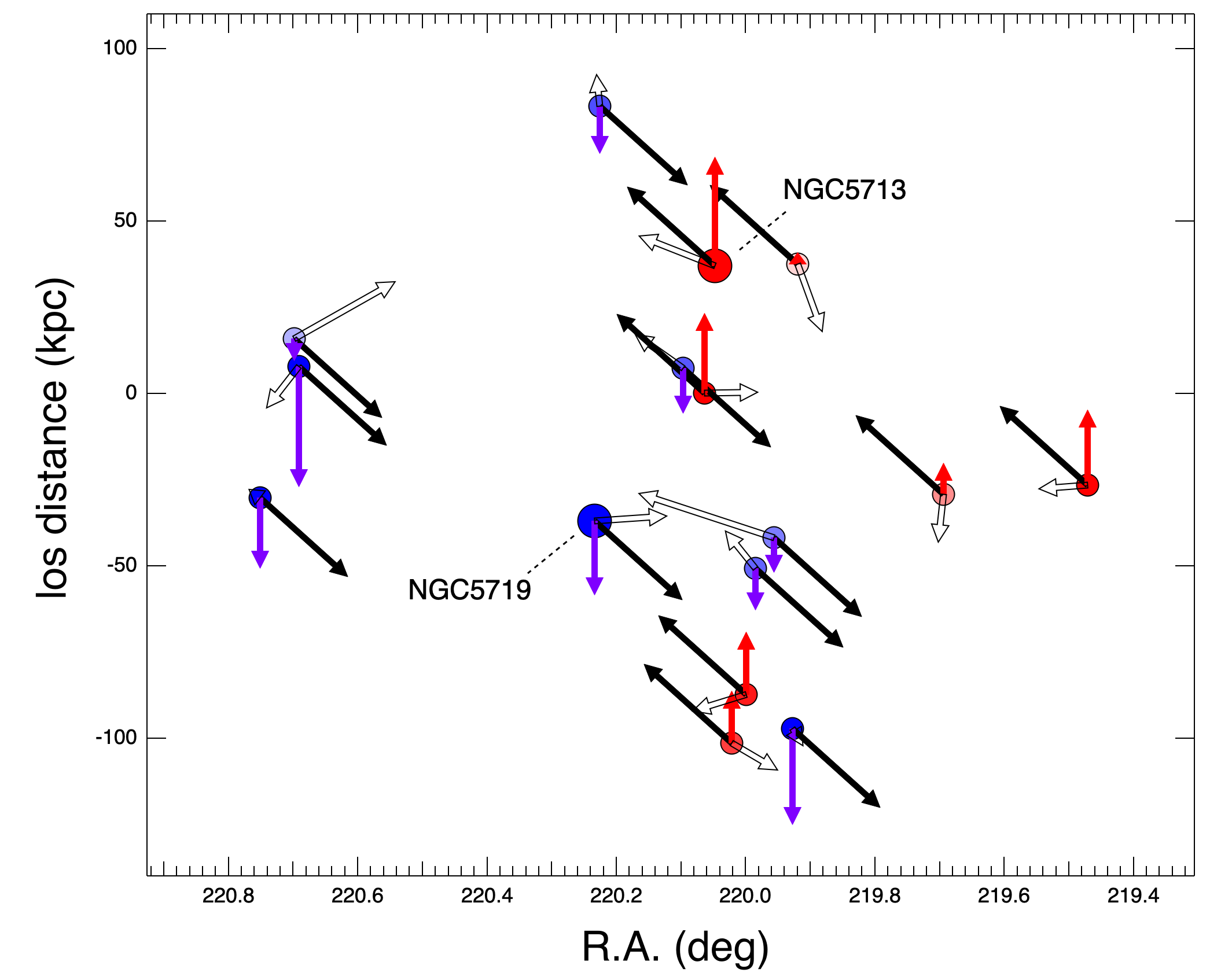}
    \caption{An example of an infall configuration that produces the observed heliocentric velocities. NGC5719 with satellites is moving in from the east at an angle of $-41\deg$, while  NGC5713 with satellites is moving in from the west at an angle of $+41\deg$. The R.A. coordinates are the observed values, the line-of-sight distances ($y$-axis) are unknown and were randomly generated in the range $\pm 150$\,kpc, except for NGC5713 ($d=35$\,kpc) and NGC5719 ($d=-35$\,kpc). The three types of velocity arrows are: (white arrow) velocity from the tidal interaction, (black arrow) bulk flow of the galaxy groups representative for the Bo{\"o}tes Sheet distance gradient, (blue/red arrows) line-of-sight component of the sum of interaction and bulk velocity vectors as observed.}
  \label{infall_scenario}
\end{figure}

As the angle $\phi$ is random there are many solutions. We show a schematic of one possible infall configuration for the NGC5713 and NGC5719 galaxies with their satellites that leads to the observed heliocentric velocities in Fig.\ref{infall_scenario}. The R.A. coordinates are the observed values, the line-of-sight distances of the galaxies ($y$-axis) are unknown and were randomly selected in the range $-150<d<150$\,kpc, except for NGC5713 and NGC5719 which were placed at $d=35$\,kpc and $d=-35$\,kpc, respectively. The three types of velocity vectors are: (white arrow) velocity of the satellite around new centre of mass, (black arrow) infall velocity of the galaxy groups representative for the Bo{\"o}tes Strip distance gradient, (blue/red) line-of-sight component of the the sum of the two velocity vectors. Given NGC5713 and NGC5719 have almost identical masses we can compare our results with simulations of equal-mass spiral mergers \citep{Lotz2008, Ji2014}
and more specifically with hydrodynamic simulations of the 
future Milky Way and Andromeda interaction \citep{cox2008}. The estimated relative spatial velocity of 204\kms between NGC5713 and NGC5719 and their separation suggest that the merger is 2-2.5\,Gyr more advanced than the Local Group, as per Figs.\,6 and 7 in \cite{cox2008}.

\section{NGC5713/19 group analogues in Illustris-TNG}
 To investigate how common the observed phase-space correlation of two merging  satellite systems is in cosmological $\Lambda$CDM simulations we have used the Illustris-TNG100\footnote{https://www.tng-project.org/} suite of hydrodynamical simulations of galaxy formation \citep{Nelson2019} adopting $\Omega_M=0.3$, $\Omega_\Lambda=0.7$, $h=0.7$, and searched for analogs of the NGC5713/19 group. The selection criteria were:
\begin{itemize}
\item two host galaxies with stellar masses: $10<\log(M_*/M_\odot)<11$.
\item host galaxy spatial separation: $50-200$\,kpc.
\item 10-20 satellites with $7.5<\log(M_*/M_\odot)<9.5$, {\tt SubhaloFlag $\neq$ 0}, and within the nominal group's virial radius, where the upper and lower limits reflect the stellar masses of the brightest (NGC5691) and the faintest NGC5713/19 group satellites (AGC540925, PGC214314), respectively. The equivalent median number of stellar particles that this lower mass limit translated to in TNG100 is 64.
\item no other, more massive satellites within the virial radius.
\end{itemize}

\begin{figure}
\includegraphics[width=1\columnwidth]{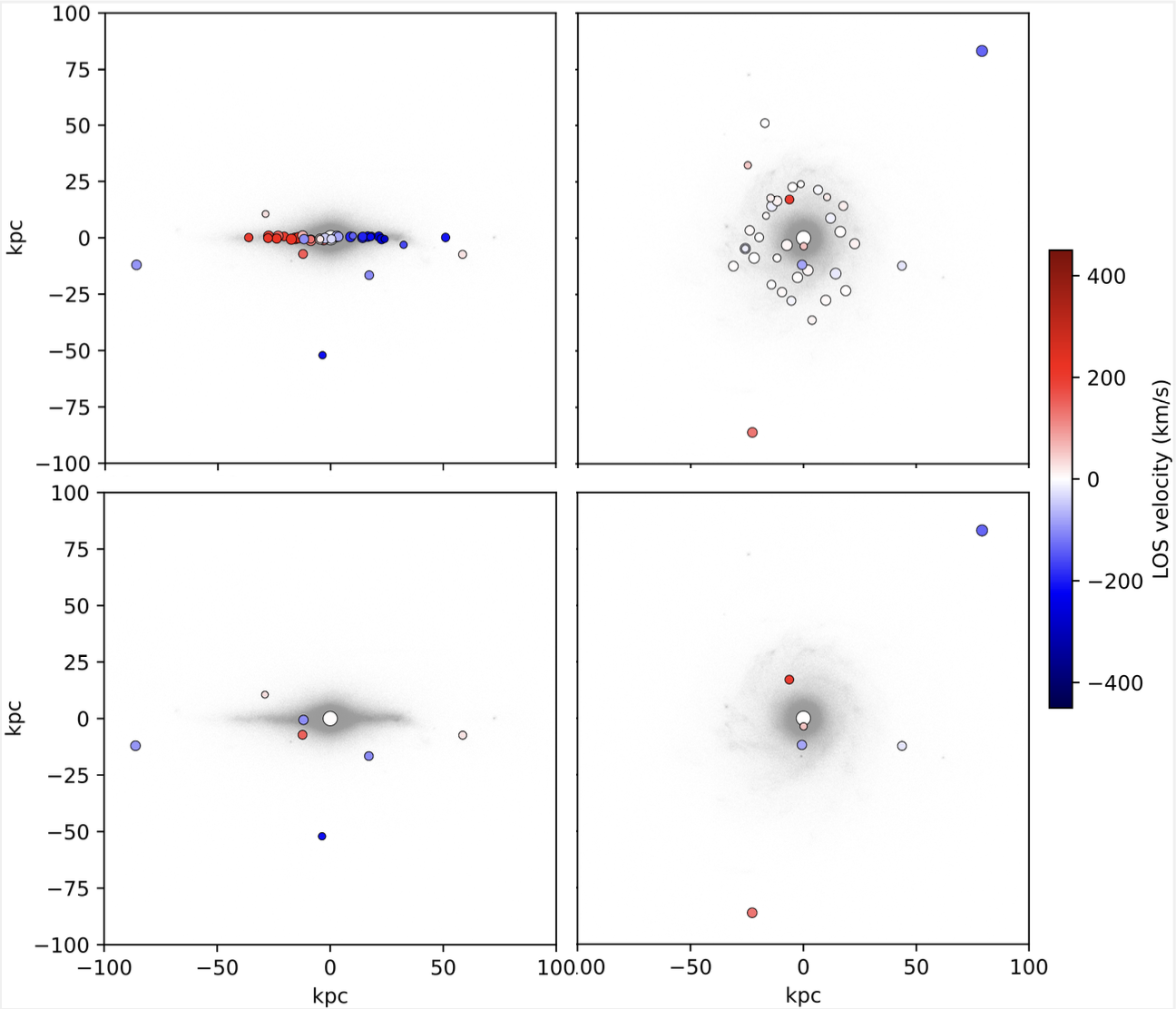}
\caption{Illustration of spurious objects mimicking an almost perfect co-rotating flattened structure of subhalos in the IllustrisTNG Group 81. Top row: edge and face-on view showing the positions and LOS velocities of all 36 subhalos relative to the host galaxy (circled object at the origin with stellar density in grey), including the 29 subhalos flagged as spurious objects ({\tt SubhaloFlag = 0}). Symbol sizes are proportional to the halo mass. According to TNG these objects should not be considered to be satellite galaxies. Bottom row: Once the spurious satellites have been removed from the group sample there is no evidence of a significant satellite co-rotation or flattening left.}
\label{spurioushalos}
\end{figure}

We like to briefly highlight the importance of excluding spurious satellites (objects with {\tt SubhaloFlag = 0}) in such an analysis using Group 81 within Illustris-TNG50 as an example. This group (Halo id: 81) was identified in  \cite{Hu2025} as featuring a very tight rotating plane of satellites, however we show that this conclusion is solely based on spurious subhalos. The top row of Fig.\ref{spurioushalos} shows the positions and relative LOS velocities of 39 (out of 41) subhalos in this group viewed edge-on and face-on, along with the host galaxy's stars shown in grey. The majority of these halos lie within the outer disk of the host galaxy, suggesting that they are in fact clumps within the disk which have been misidentified as individual halos. The raw sample indeed includes 31 subhalos flagged as spurious objects ({\tt SubhaloFlag = 0}). Taken at face value the satellite system shows an almost perfect co-rotating flattened structure. However, the Illustris team warns against using spurious objects. They should not be considered to be satellite galaxies. Once the spurious satellites have been removed from the group there is no evidence of a significant satellite co-rotation or flattening left, as seen in the bottom row of Fig.\ref{spurioushalos}.

\begin{table}
\caption{Groups in the TNG100 simulation matching our criteria for NGC5713/19 analogues. (1) ID of the group within the FoF Halo catalogue. (2,3) Stellar masses of the two host galaxies of the group. (4) Separation between the two hosts. (5) Number of dwarf galaxies ($7.5<\log(M_*/M_\odot)<9.5$) in the group. (6) Relative velocity between the two host galaxies. (7) Fraction $f$ of dwarf galaxies co-rotating with the host pair. }
\begin{tabular}{ccccccc}
\hline
Group &  Host 1 & Host 2 &  Separation & $N_t$& $\delta v$ & $f$\\
ID & log(M$_{\odot}$) & log(M$_{\odot}$) & (kpc) & & (km/s) & \\
(1) &  (2) & (3) &  (4) & (5) & (6) & (7) \\

 \hline 
363 & 10.5 & 10.4 & 134 & 11 & -189 & 0.91\\
430 & 10.4 & 10.3 & 79 & 11 & -164 & 0.82\\
312 & 10.8 & 10.6 & 97 & 12 & -105 & 0.75\\
347 & 10.7 & 10.2 & 42 & 13 & 44 & 0.69\\
417 & 10.6 & 10.5 & 164 & 11 & 71 & 0.64\\
268 & 10.8 & 10.5 & 99 & 13 & -3 & 0.62\\
292 & 10.7 & 10.7 & 161 & 15 & -25 & 0.53\\
294 & 10.6 & 10.7 & 95 & 12 & -40 & 0.50\\
\hline
\end{tabular}
\label{TNG_groups}
\end{table}

Returning to our search for NGC5713/19 analogues, after removing spurious halos and applying our selection criteria detailed above,   we found a total of eight galaxy groups within the 
$110.7^3$ Mpc$^{3}$
volume of the simulation. The host masses, host separation and the number of dwarf galaxies are given in Table~\ref{TNG_groups}, along with their group ID (i.e. their index) in the fof Halo catalogue for TNG100. The small sample size of NGC5713/19 analogues from the volume of the TNG100 simulation is limited by the combined requirements for close host separation and for the number of luminous satellites. Relaxing the host separation limit to 500\,kpc results in a doubling of the sample to 16, while removing the lower limit for the number of luminous satellites (and maintaining a maximum host separation of 200\,kpc) significantly increases the sample to 63 groups. This highlights that the combination of a large number of luminous satellites and small host separation is relatively uncommon, with the number of satellites above $\log(M_*/M_\odot)=7.5$ being the most restricting criterion.

For each group, we firstly determined the total angular momentum vector of the two high-mass galaxies about their mutual centre-of-mass and re-orientated the system such that this vector is directed along the $z$-axis. We then further rotated the system such that the line connecting the two host galaxies lies along the $x$-axis, i.e. in the plane of the sky, and used the $y$-component of each galaxy's velocity as its line-of-sight velocity. Using the host angular momentum vector as a dividing line and noting the motion (approaching or receding) of the host on each side of this line, we determined the number of satellites which are moving in the same sense as the high-mass galaxy ($ N_{r}$) on their corresponding side. Dividing by the total number of satellites ($N_{t}$), we determined the fraction of satellites which are co-rotating with the high-mass galaxy pair
($f=N_{r}/N_{t}$). 

\begin{figure}
\includegraphics[width=1\columnwidth]{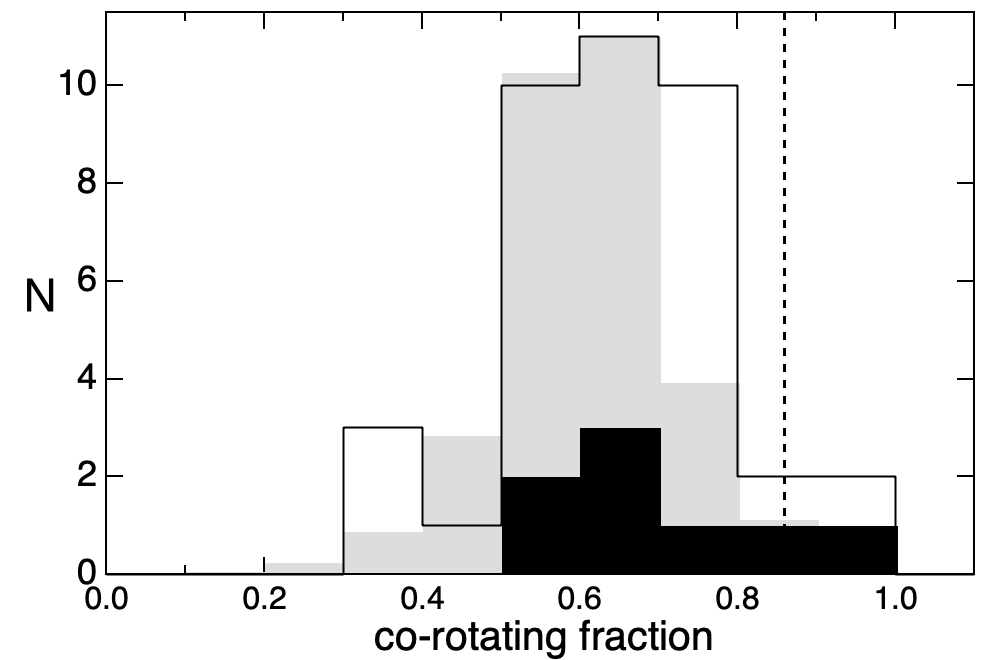}
\caption{Distribution of the fraction of satellites co-rotating with the two hosts within TNG100 analogues of the NGC5713/19 group. The filled histogram shows the distribution for the eight groups matching the criteria given in the text, showing a preference towards co-rotation. The unfilled histogram shows that this preference holds when the lower mass limit of satellites is reduced from $10^{7.5}$ to $10^{6.5} M_{\odot}$, allowing for a larger galaxy group sample. The vertical dashed line shows the co-rotating fraction $f=12/14=0.86$ for the NGC5713/19 system. For comparison we also show the $f$ distribution for satellites in isotropy (grey histogram) derived from 5000 satellite systems (10-20 satellites each) with random velocities and positions in a spherical halo around a single host. That histogram is scaled to match the peak of the histogram for the 39 lower-mass Illustris-TNG satellite systems.}
\label{corotating_fraction}
\end{figure}
\begin{figure}
\includegraphics[width=1\columnwidth]{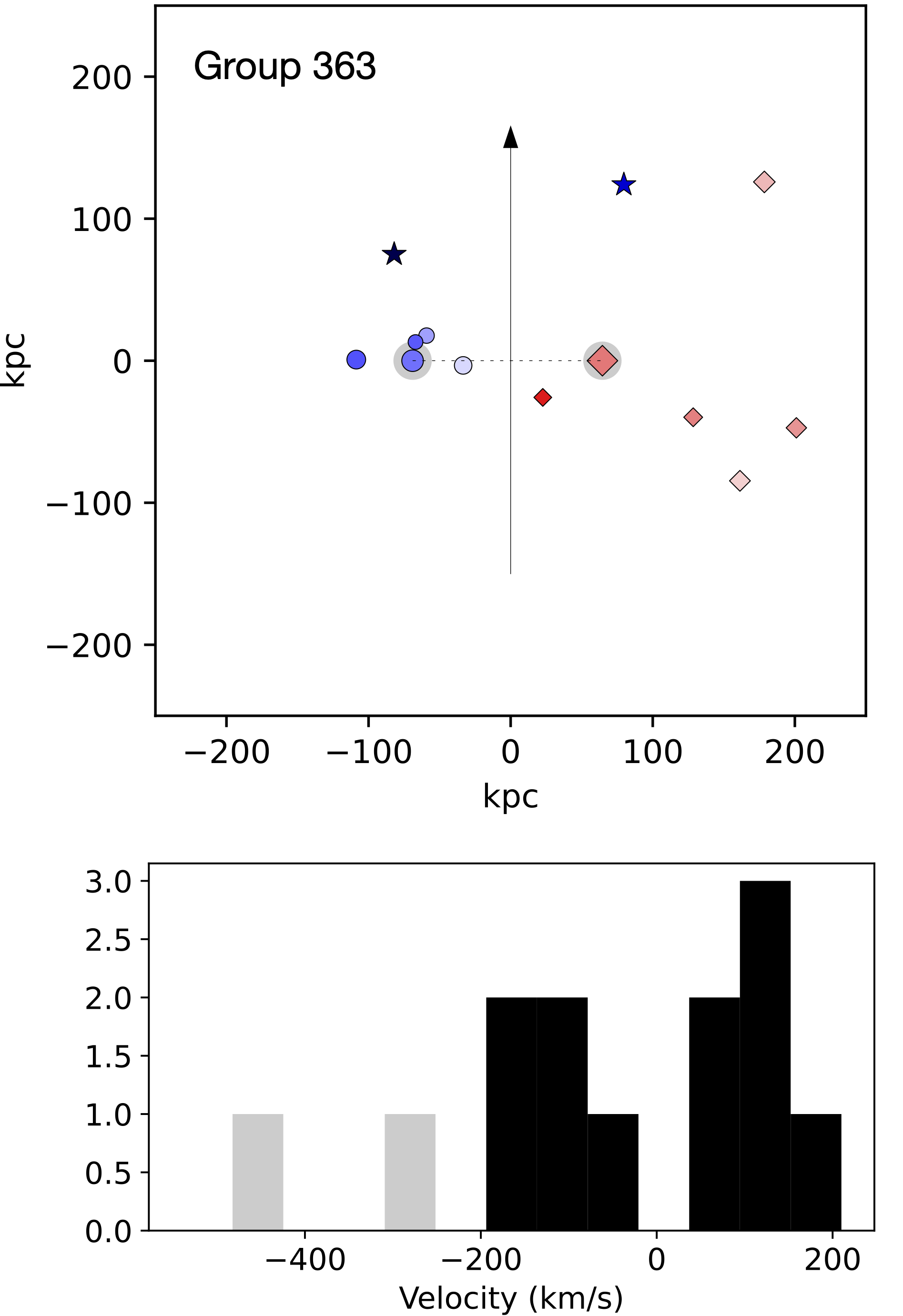}
\caption{Groups 363 is the NGC5713/19 group analogue from TNG100 which shows the greatest degree of satellite coherent motion about the angular momentum vector of the two hosts. The top panel shows the projected spatial distribution and relative LOS velocities of the members of the system, aligned such that the angular momentum vector of the host (black arrow) is in the plane of the projection and oriented vertically. Additionally, the system is rotated such that the connecting line between the host lies in the plane, along the $x$-axis. The hosts are highlighted by a grey outline and are presented either as a circle or diamond. Satellites which are associated with one of the hosts prior to infall (i.e. have entered an orbit around the host, see Figure~\ref{group_history}) are given the corresponding symbol, while those two which do not have a strong past association with the host are shown by star symbols. The bottom panel shows the LOS relative velocity distribution, with non-associated satellites shown in grey. We note that Group 363 closely resembles the NGC5713/19 group with a bimodal LOS velocity distribution. However, the velocity range for Group 363 is about twice that of NGC5713/19 (400\kms versus 200\kms$/\sin(41^\circ)=300$\kms). 
}
\label{simulationgroups}
\end{figure}

\begin{figure}
\includegraphics[width=1\columnwidth]{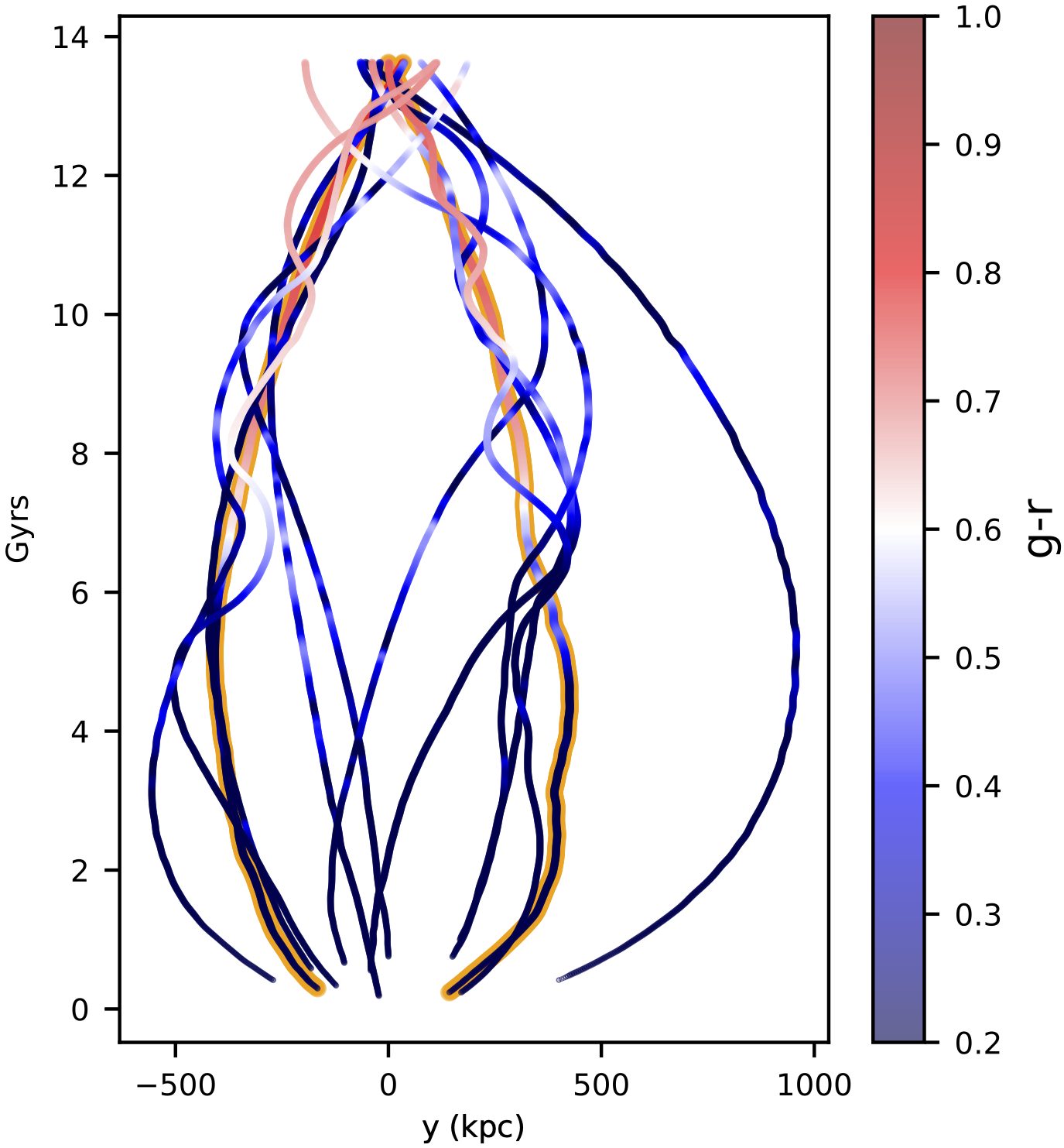}
\caption{Formation history of Group 363 within the Illustris TNG100 simulation, showing the position (projected onto the y-axis) versus time for all group members with present-day masses above $\log(M/M_{\odot})=7.5$. Galaxies are coloured by their $g-r$ colour index and the two hosts are highlighted by orange outlines. The 100 snapshots of the simulation have been interpolated to create smooth tracks. The two host galaxies of Group 363 initially formed their own systems of dwarf galaxy satellites before coming together to form the present-day group. Nine of the group's dwarfs were associated with a host prior to their first infall, while two had only made a single approach past one of the hosts.}
\label{group_history}
\end{figure}

Fig.~\ref{corotating_fraction} displays the distribution of co-rotating fractions $f$ for the eight galaxy groups (black histogram). The mean of the distribution is $\langle f\rangle=0.68\pm0.05$, indicating a preference for co-rotation. In order to test if this preference holds for a larger number of systems with two proximal hosts, the lower-stellar mass limit for the satellites was decreased from $10^{7.5}$ to $10^{6.5} M_{\odot}$, which increased the total number of NGC5713/19 look-alike groups to 39. This results in a similar shift towards co-rotation, with a mean fraction of $\langle f\rangle=0.63\pm0.02$, further supporting the preference for satellite co-rotation in galaxy groups with two high-mass galaxies in close proximity. For comparison we also show the $f$ distribution for satellites in isotropy (grey histogram). It was derived from 5000 realisations of 10-20 satellites with random velocities and positions in a spherical halo around a single host. We note that the distribution of the co-rotation fraction changes with the number of satellites in the system even under the isotropic null-hypothesis.

It is interesting to point out that these  statistical results are based on a 100 percent detection rate of such merging host galaxies with their satellites as we have access to full phase-space information. The conclusion is that only a few of these merging groups exist in Illustris-TNG100 \citep[see also][]{Kanehisa2023b}. They would be even rarer to find in reality as it requires that they can be observed under the ideal viewing angle.

The galaxy pair with the highest degree of co-rotation among its satellite members is Group 363. This group has a host separation of 134\,kpc and includes 11 satellite galaxies above $\log (M/M_{\odot})=7.5$. Fig.\ref{simulationgroups} displays the spatial distribution of the group members along with their relative LOS velocities, with the angular momentum vector of the two hosts indicated by the vertical arrow. This clearly illustrates the co-rotation of satellite galaxies with the host pair, with the majority of satellites having the same motion about the central angular momentum vector as their hosts.  
Fig.\ref{group_history} shows the formation history of this galaxy group from the Big Bang to the present day, tracing the paths of the two host galaxies and all satellite galaxies which form the present-day group. Note that galaxies which have merged into one of the other members before the present day are not shown in this figure. The two hosts were initially well separated from each other and formed their own systems of satellite galaxies, which can be seen here orbiting around their respective host, before the two hosts came together and formed the present-day group. Looking at the bigger picture, we see that Group\,363 is located near the intersection of two cosmic filaments, and the tracks of the two hosts suggests each host-and-satellites traveled along a separate filament. We associate a satellite to one of the hosts if it has entered into an orbit around the host (seen here by a satellite making at least two crossings with a host), or as an unassociated satellite if it did not fall into an orbit before the two hosts came together. In Fig.\ref{simulationgroups}, host-associated galaxies are shown with the same symbol as their respective host, or as a star if they were not previously associated with a host. Of the 10 satellites which share the same motion as the host on their respective side of the angular momentum vector, nine are associated with that host, showing that the observed co-rotation is due to the motions of the two host-satellite systems as they come together. The remaining co-rotating satellite, along with the only counter-rotating satellite, fell into the group from a larger distance and were not previously associated with either host. It is likely that similar formation scenarios for the other groups in the simulation has led to the predominance of co-rotation among the two host galaxies and their satellites.

\begin{table*}
\caption{Satellite galaxies within the virial radius of the host galaxy pair NGC5719/13. (1) galaxy name; (2) GAMA Survey ID, (3) right ascension in epoch J2000; (4) declination in epoch J2000; (5) heliocentric velocity; (6) velocity relative to the group's systemic velocity of 1805\kms, (7)  distance from the rotation axis along the major axis (Section 3.1), (8) total $K_s$-band luminosity derived from $K_s$-band magnitude in \protect\cite{Kourkchi2017}
, except for AGC540925 and PGC214314, (9) proposed kinematic membership for the edge-on scenario (Section 4.2) - CR: co-rotating, ISO: isotropic, (10) reference for the velocity measurement.}
\begin{tabular}{lrcccccccl}
\hline
Name &  GAMA & R.A.& DEC & $v_\odot$ & $\Delta v_\odot$  & $\Delta R_{proj}$ & $\log(L_{\rm K})$ & MEM & Reference\\
& & (deg) & (deg) & (km\,s$^{-1})$ & (km\,s$^{-1})$   & (kpc) &  ($\log L_{\rm K_\odot}$) &  (CR/ISO) &\\
         (1) &          (2) &          (3) &            (4) &           (5) & (6) & (7) & (8)& (9) & (10)\\
 \hline
 NGC5691                 & 64553   & 219.4715   & $-0.3993$   & 1870$\pm 8$ & 65  & 330.5 & 10.15 & CR & \cite{Theureau1998}\\
AGC540925             & 594302 & 219.6944 & $-0.0271$ & 1832$\pm 3$ & 27 & 257.0 & $<7.78$ & CR  & \cite{Haynes2018} (Aug2019)\\
PGC135857              & 64758  & 219.9195   & $-0.3032$   & 1815$\pm 8$ & 10    & 108.3 & 8.30&  CR  & \cite{DESI2024}\\
PGC1155970             & 79084  & 219.9276   & $0.0560$     & 1722$\pm8$ & $-83$  & 204.8 & 7.93& ISO  & \cite{DESI2024}\\
 NGC5705                & 49167  & 219.9569   & $-0.7199$   & 1775$\pm 11$ & $-30$  & 222.1 & 9.48& ISO  & SDSS-DR13 \citep{SDSSDR13}\\
PGC214313  & 49173  & 219.9849   & $-0.7443$   & 1769$\pm 64$ & $-36$  & 228.1 & 9.18&  ISO  & 2dFRS \citep{Colless2001}\\
PGC135858            & 594420 & 219.9996 & $-0.1863$ & 1859$\pm 10$ & 54   & 90.1& 7.82&  CR   & \cite{Roberts2004}\\
PGC214314              & 49204   & 220.0216   & $-0.7226$   & 1850$\pm 8$ & 45   & 212.6 & $<7.78$ & ISO   & \cite{DESI2024}\\
NGC5713                  & 64771  & 220.0479   & $-0.2901$   & 1899$\pm 8$  & 94   & 46.0 & 10.78 & CR   & \cite{Vergani2007}\\
PGC1159795             & 79098  & 220.0635   & $0.2069$    & 1874$\pm 3$ & 69   & 252.9 & 8.63& ISO  & \cite{Haynes2018}\\
PGC1140314 & 569709 & 220.0969   & $-0.5622$   & 1766$\pm15$ & $-39$  & 127.8 & 8.80 &  CR  & \cite{DESI2024}\\
 GAMA64829                & 64829  & 220.2256   & $-0.3557$   & 1764$\pm 5$ & $-41$  & 48.4  & 9.48& CR & GAMA DR4 \citep{Driver2022}\\
NGC5719                  & 64804  & 220.2340    & $-0.3189$   & 1741$\pm 8$ & $-64$  &  46.0  & 10.86 & CR  & \cite{Vergani2007}\\
NGC5733                  & 64893  & 220.6916   & $-0.3507$   & 1701$\pm 8$ & $-104$ & 270.1 & 9.53& CR  & \cite{DESI2024}\\
PGC1141860             & 569842 & 220.6990    & $-0.4985$   & 1786$\pm  8$ & $-19$  & 288.8 & 8.08& CR & \cite{DESI2024}\\
PGC135860              & 64937  & 220.7516   & $-0.3835$   & 1744$\pm 20$ & $-61$  & 301.0 & 8.73 & CR & \cite{impey1996} \\
 \hline
\end{tabular}
\begin{tablenotes} 
\item Note: the interacting galaxies NGC5705 and PGC214313 are considered to be a single object for the mixed satellite kinematics analysis. 
\end{tablenotes}
\label{galaxyparameters}
\end{table*}

\section{summary and conclusions}
The merging galaxy pair NGC5713/19 was found to have a satellite system with a distinct bimodal distribution in terms of sky positions and relative line-of-sight velocities, with all but one satellite south east of NGC5719 being blueshifted and all but one satellite north west of NGC5713 being reshifted. They also follow the same motion as their nearby host galaxies. 

Using the centre of mass of the two host galaxies and the on-sky separation of redshifted and blueshifted galaxies produces a position-velocity diagram with an S-shaped phase-space correlation that resembles a flat rotation curve with an amplitude of $67\pm12$\kms. The intergalactic HI gas associated with the galaxy pair interaction follows the same trend but with a higher amplitude.

We determined that the observed galaxy velocity field is unlikely the result of galaxies moving isotropically in a combined NFW dark matter halo.
We investigated three alternative scenarios that also consider additional information we have for the NGC5713/19 environment.

\noindent {\bf Nearly face-on co-rotating plane of satellites}, which would explain the observed
low ellipticity of the 2D satellite distribution ($b/a=0.752\pm0.216$) and the low maximum
line-of-sight rotation velocity ($67\pm12$\kms). Adopting the same intrinsic plane thickness as observed for MW and M31 we derived a plane inclination angle of $i_{plane}=42.3\degr$ and a true satellite rotation velocity of $\bar{v}_{max}=99.6\pm17.8$\kms. The observed galaxies have a Spearman rank order coefficient of $r_s=0.46$, a value that becomes comparable to the rank order coefficients for modeled galaxies in a rotating disc embedded in an external NFW halo if the disc is kinematically hot, i.e. the galaxy random motions are of order of the circular velocity.

\noindent {\bf Kinematically mixed satellite system made up of a co-rotating plane seen almost edge-on and an isotropic component}. This would explain the low ellipticity of the 2D satellite distribution, the presence of some counter-rotating satellites and it would be consistent with the interaction of NGC5713/19 seen edge-on. We allocated the two counter-rotating satellites to the isotropic component and selected two more satellites to achieve the highest flattening among the remaining co-rotating satellites.  We found an inclination angle of $i_{plane}=68\degr$ for the satellite plane and a true satellite rotation velocity of $\bar{v}_{max}=71\pm13$\kms. The position angle of the satellite plane rotation axis ($PA=6.7\degr$) is found to be almost perpendicular ($87.9\degr$) to the central HI gas bridge and to the connection line between the host galaxies NGC5713 and NGC5719. 

\noindent {\bf Merger between two host galaxies with their satellites}. This scenario is inspired by the observations that the central object in the group is not a single host but a galaxy pair in the early stage of merging, and the plane of interaction of NGC5713/19 is seen close to edge-on as judged from the linear morphology of the associated HI bridge. It explains not only the galaxy group properties such as sky separation of blue and reshifted galaxies, low ellipticity in the on-sky distribution, satellites moving in the same sense as their nearby parent galaxies, but is also fully consistent with the implications of NGC5713/19 and their satellites being embedded in the Bo{\"o}tes Strip cosmic structure. The Bo{\"o}tes Strip has a well-defined distance gradient of $-0.42$\,Mpc/degree \citep{Karachentsev2014} at the location of NGC5713/19, 
meaning that Bo{\"o}tes Strip galaxies on the eastern side of NGC5713 are systematically closer to us than galaxies on the NGC5719 side.
The distance gradient predicts that NGC5719 with its entourage of satellites was moving to the current location from behind and NGC5719 from the front. Hence the LOS velocities of the NGC5713/19 satellites are dominated by infall. We estimates the bulk flow infall velocities for both hosts to be $v_{infall}=102\pm18$\kms as they have approximately the same mass. The estimated relative
velocity of 204\kms between NGC5713 and NGC5719 and their
separation suggest that this galaxy merger is 2-2.5 Gyr more advanced than the Local Group, 

All three scenarios are plausible to explain the galaxy velocity field around NGC5713/19 with the significant difference that the merger scenario is the only one that also takes into account and is consistent with the fact that the GAMAJ1440-00 group is undergoing a galaxy group merger and is embedded in a larger scale cosmic structure with specific properties. 
The case of the GAMAJ1440-00 group demonstrates that a satellite system around an interacting galaxy pair can produce a line-of-sight velocity field that mimics a co-rotating satellite system.

To test the number of galaxy groups with merging central $L^*$ host galaxies in simulations we search for NGC5713/19 analogies in IllustrisTNG-100. A small number of only eight galaxy groups have similar properties. Their mean satellite co-rotation fraction is $\langle f\rangle=0.68\pm0.05$. A slightly relaxed constraint on the lower mass limit for the satellites yields a larger group sample of 39 with a similar result of $\langle f\rangle=0.63\pm0.02$.
The group with the highest co-rotating satellite fraction (Group 363) happens to have also a similar bimodal LOS velocity distribution to NGC5713/19 and their satellites.

We also report eight new LSB dwarf satellite candidates within the virial radius of NGC5713/19, three more star-formation clumps around NGC5713/19 and a previously unnoticed stellar stream west of NGC5713.

Based on our findings we can make a number of recommendations for follow-up studies: 

An additional 18  low surface brightness satellite candidates without velocities are known in the GAMAJ1440-00 group. Basic properties are listed in Table\,\ref{candidates}. Redshift measurements for them are desirable to confirm their group memberships and further test the statistical significance of the coherent velocity field around NGC5713/19.
In this context it would be also interesting to search and spectroscopically follow up globular cluster candidates to determine the central velocity field in the immediate vicinity of the pair.

The closest satellite to NGC5713 is PGC135858. Measuring the chemical abundances for it at the far end of the HI tidal tail and compare them with chemical abundances of the mass-metallicity relation could help to confirm if it is a tidal dwarf following the work by \citet{Sweet2014}.

The combined satellite system around NGC5713/19 is expected to remain a co-rotating structure once the host galaxies fully merged and formed an early-type galaxy. Possibly what we observe around Centaurus\,A \citep{Tully2015, Muller2016,Muller2018a, Muller2021} and NGC4490/85 \citep{Karachentsev2024, Pawlowski2024}. That raises the question about the frequency of co-rotating post-merger satellite systems.
The MATLAS results \citep{Heesters2021} have already shown statistically the presence of 25\% flattened satellite systems around hosts with early-type morphology. Those groups with their satellites would be prime targets for spectroscopic follow-up to systematically search for evidence of co-rotation among the satellites.

\newpage

\appendix
\section{Satellite candidates }
An additional 18  low surface brightness satellite candidates without velocities are known in the GAMAJ1440-00 group ($219.35\deg<R.A.<220.95\deg$, $-1.13\deg<DEC<0.49\deg$). They are dwarf galaxies with suspected
group membership based on morphology and angular size, but without velocity measurements. A detailed analysis and discussion of the candidates 
will be provided in an upcoming paper. For completeness we list 
basic parameters in Table\,\ref{candidates}.

\begin{table}
\caption{Satellite candidates and other features within the vicinity ($219.35\deg<R.A.<220.95\deg$, $-1.13\deg<DEC<0.49\deg$) of the host galaxy pair NGC5719/13. (1) galaxy name (2)  right ascension in epoch J2000; (3) declination in epoch J2000; (4) reference.}
\begin{tabular}{lccl}
\hline
Name &  R.A. & DEC  &  reference\\
& (deg) & (deg) & \\
(1) & (2) & (3) & (4) \\
 \hline
{[GGS2018]} 597 & 219.4326 & 0.2697 & \cite{Greco2018}\\
GAMA79027 (MGC65479)	& 219.5770	& 0.2066 & \cite{Liske2003} \\ 
{[GGS2018]}  690 & 	219.6894	& $-0.7057$ & \cite{Greco2018}\\
{[GGS2018]}  590	& 219.9292	& $-$0.2772 & \cite{Greco2018} \\ 
{[GGS2018]}  589	& 219.9618	& 0.3531 & \cite{Greco2018} \\ 
{[GGS2018]}  684 & 220.0278 & $-1.0209$  & \cite{Greco2018}\\
dw220.07$-$0.24	& 220.0657	& $-0.2441$ & this study \\ 
dw220.12$-$0.37	& 220.1193	& $-0.3743$ & this study \\ 
{[GGS2018]}  685 & 220.1928 & $-0.7347$ & \cite{Greco2018}\\
dw220.29$-$0.33	& 220.2942	& $-0.3279$ & this study \\ 
{[GGS2018]} 688	& 220.3196	& $-0.5930$ & \cite{Greco2018} \\ 
{[GGS2018]} 686	& 220.4287	& $-0.6577$ & \cite{Greco2018} \\ 
dw220.49$-$0.33	& 220.4875	& $-0.3313$ & this study \\ 
dw220.53$-$0.77	& 220.5349	& $-0.7679$ & this study \\ 
dw220.68$-$0.32	& 220.6845	& $-0.3218$ & this study \\ 
dw220.69$-$0.47	& 220.6889	& $-0.4735$ & this study \\ 
dw220.70$+$0.27	& 220.6986	& 0.2662 & this study \\ 
{[GGS2018]} 591	& 220.7370	& 0.2802 & \cite{Greco2018}  \\ 
 \hline
 \multirow{2}{10em}{Western Stellar Stream} & 219.98	 & $-0.3258$ & \multirow{2}{4em}{this study} \\ 
& 220.00 & $-0.3075$  & \\ 
Clump-5 & 219.9799	& $-0.3238$ & this study \\ 
 Clump-4	& 220.1568	& $-0.2978$ & this study \\ 
Clump-3	& 220.2156	& $-0.3000$ & this study 
\\ 
Clump-2	& 220.2588	& $-0.3330$ & \cite{Neff2005} \\ 
 Clump-1 (GAMA64831)	& 220.2875	& $-0.3546$ & \cite{Neff2005} \\ 
\hline
\end{tabular}
\label{candidates}
\end{table}

\section*{Acknowledgments}
The authors thank Marcel Pawlowski, Oliver M\"uller, and the anonymous referee for feedback and a constructive report, which helped to clarify and improve the manuscript.
HJ thanks the Queensland University School of Mathematics and Physics for hospitality and funding support as part of their distinguished visitor program. We thank Khaled Said Soliman and Cullan Howlett from the DESI team for providing information to clarify the case of GAMA594420. SMS acknowledges funding from the Australian Research Council (DE220100003).
Parts of this research were conducted by the Australian Research Council Centre of Excellence for All Sky Astrophysics in 3 Dimensions (ASTRO 3D), through project number CE170100013.
This research used data obtained with the Dark Energy Spectroscopic Instrument (DESI). DESI construction and operations is managed by the Lawrence Berkeley National Laboratory. This material is based upon work supported by the U.S. Department of Energy, Office of Science, Office of High-Energy Physics, under Contract No. DE-AC02-05CH11231, and by the National Energy Research Scientific Computing Center, a DOE Office of Science User Facility under the same contract. Additional support for DESI was provided by the U.S. National Science Foundation (NSF), Division of Astronomical Sciences under Contract No. AST-0950945 to the NSF's National Optical-Infrared Astronomy Research Laboratory; the Science and Technology Facilities Council of the United Kingdom; the Gordon and Betty Moore Foundation; the Heising-Simons Foundation; the French Alternative Energies and Atomic Energy Commission (CEA); the National Council of Science and Technology of Mexico (CONACYT); the Ministry of Science and Innovation of Spain (MICINN), and by the DESI Member Institutions: www.desi.lbl.gov/collaborating-institutions. The DESI collaboration is honored to be permitted to conduct scientific research on Iolkam Du'ag (Kitt Peak), a mountain with particular significance to the Tohono O'odham Nation. Any opinions, findings, and conclusions or recommendations expressed in this material are those of the author(s) and do not necessarily reflect the views of the U.S. National Science Foundation, the U.S. Department of Energy, or any of the listed funding agencies.

\section*{Data Availability}
The data on which this work is based is publicly available and is detailed in the corresponding sections of this manuscript.

\bibliographystyle{mnras}
\bibliography{corotation.bib}

\bsp
\label{lastpage}
\end{document}